\documentclass[
reprint,
amsmath,amssymb,
aps]{revtex4-2}

\usepackage{gensymb}
\usepackage{xurl}
\usepackage{multirow}

\usepackage{graphicx}
\usepackage{dcolumn}
\usepackage{bm}

\begin{document}

\title{Predicting the role of inequalities on human mobility patterns}

\author{Alain Boldini}
\affiliation{Department of Mechanical Engineering, New York Institute of Technology, College of Engineering and Computer Science, Old Westbury, New York 11568, USA}
\affiliation{Center for Urban Science and Progress and Department of Mechanical and Aerospace Engineering, New York University, Tandon School of Engineering, Brooklyn, New York 11201, USA}

\author{Pietro De Lellis}
\affiliation{Department of Electrical Engineering and Information Technology, University of Naples Federico II, Naples, NA 80125, Italy}

\author{Salvatore Imperatore}
\affiliation{Department of Electrical Engineering and Information Technology, University of Naples Federico II, Naples, NA 80125, Italy}

\author{Rishita Das}
\affiliation{Center for Urban Science and Progress and Department of Mechanical and Aerospace Engineering, New York University, Tandon School of Engineering, Brooklyn, New York 11201, USA}
\affiliation{Department of Aerospace Engineering, Indian Institute of Science, Bengaluru, Karnataka 560012, India}

\author{Luis Ceferino}
\affiliation{Center for Urban Science and Progress and Department of Civil and Urban Engineering, New York University, Tandon School of Engineering, Brooklyn, New York 11201, USA}
\affiliation{Department of Civil and Environmental Engineering, College of Engineering, University of California Berkeley, Berkeley, California 94720, USA}

\author{Manuel Heitor}
\affiliation{Center for Urban Science and Progress and Marron Institute of Urban Management, New York University, Tandon School of Engineering, Brooklyn, New York 11201, USA}
\affiliation{Center for Innovation, Technology and Policy Research, Instituto Superior Técnico, Technical University of Lisbon, Lisbon, 1049-001, Portugal}

\author{Maurizio Porfiri}
\thanks{Author for correspondence}
\email{mporfiri@nyu.edu}
\affiliation{Center for Urban Science and Progress, Department of Mechanical and Aerospace Engineering, and Department of Biomedical Engineering, New York University, Tandon School of Engineering, Brooklyn, New York 11201, USA}
\date{\today}

\begin{abstract}
Whether in search of better trade opportunities or escaping wars, humans have always been on the move. For almost a century, mathematical models of human mobility have been instrumental in the quantification of commuting patterns and migratory fluxes. Equity is a common premise of most of these mathematical models, such that living conditions and job opportunities are assumed to be equivalent across cities. Growing inequalities in modern urban economy and pressing effects of climate change significantly strain this premise. Here, we propose a mobility model that is aware of inequalities across cities in terms of living conditions and job opportunities. Comparing results with real datasets, we show that the proposed model outperforms the state-of-the-art in predicting migration patterns in South Sudan and commuting fluxes in the United States. This model paves the way to critical research on resilience and sustainability of urban systems.
\end{abstract}

\maketitle

Mobility has always been part of human history \cite{mcneill1984human}. Even before the advent of \textit{Homo sapiens}, archaic humans have migrated across continents to find better survival opportunities \cite{goldstein2002human}. Migrations have been fundamental for the establishment and development of countries such as the United States (US) \cite{hansen1940atlantic}, with settlers running away from wars, religious intolerance, famines, and extreme poverty \cite{breen2016colonial,klein2012population}. In the last couple of decades, we are observing a new wave of migrations, both internal (such as from rural areas to cities \cite{zhang2003rural,parida2020migration}) and international (including the migrant crisis at the US-Mexico border \cite{US_Mexico_crisis} and the migration routes in the Mediterranean sea \cite{Mediterranean_crisis}). Migrants may not only look for better job opportunities, but also flee conflicts, natural hazards, and political persecution and instability. Human mobility is not limited to migrations: with the emergence of large metropolitan areas and mass transit systems during industrialization, people have started commuting daily from their homes in the suburbs to their workplace in city centers \cite{shen2000spatial}.

	Due to its importance in human existence and history, mobility has been approached by a variety of disciplines beyond geography \cite{lewis2021human,clark1985human}: behavioral ecology, which seeks to interpret human migrations by drawing parallels with animal mobility \cite{borgerhoff2003human,meekan2017ecology}; economics, which analyzes incentives for mobility and their effects on development \cite{sjaastad1962costs,de2010migration}; and, more recently, physics and data science, which leverage the new, large availability of mobility data to build new descriptions of commuting and migration patterns \cite{barbosa2018human,azose2019estimation}. Mathematical models of human mobility help us quantitatively describe migration and commuting patterns, supporting critical decision-making process to prepare for and manage future mobility patterns \cite{bettencourt2021introduction}. Building on substantive research in the social sciences highlighting the drivers of human mobility \cite{borjas1989economic,levitt2003international,xiang2014migration,amelina2016anthology}, several researchers attempted to mathematically predict mobility patterns at different spatio-temporal scales \cite{barbosa2018human,gonzalez2008understanding,song2010modelling,jia2012empirical,pappalardo2015returners,stouffer1940intervening,schneider1959gravity}. One of the earliest population-level description of human mobility is the gravity model \cite{ravenstein1885laws,zipf1946p},  which posits the flux from location $i$ to location $j$ as Newton's gravitational law, with populations as masses and an inverse dependence on a function $f(r_{ij})$ of the distance $r_{ij}$ between the two locations. Such a phenomenological model incorporates several intuitions regarding mobility: the number of people leaving one location should depend on its population; the attractivity (and therefore the number of people received) of a location should depend on its population; and distance should act as a modulating factor, whereby fluxes should decrease with distance between locations.

	Despite good accuracy and widespread use of gravity models \cite{erlander1990gravity,wilson1974urban,karemera2000gravity,patuelli2007network}, they suffer from several limitations. First, they fail to capture many features of empirical distributions, such as long-range mobility patterns \cite{simini2012universal,lenormand2016systematic,masucci2013gravity}. Second, they often require several parameters to fit real datasets \cite{balcan2009multiscale}, hindering their predictive value in new conditions or geographical area. Third, they may be analytically inconsistent, for example, they do not automatically limit the maximum flux from a location \cite{simini2012universal}. With respect to the latter limitation, microscopic approaches based on entropy maximization \cite{wilson1969use}, random utility maximization \cite{block1959random,sheppard1978theoretical,anas1983discrete,wang2021free}, and information friction \cite{anderson2011gravity} can help solve some analytical inconsistencies, but they do not provide functional forms for $f(r_{ij})$ \cite{simini2012universal}.

\begin{figure*}[htbp]
		\centering
		\includegraphics[width=.9\textwidth]{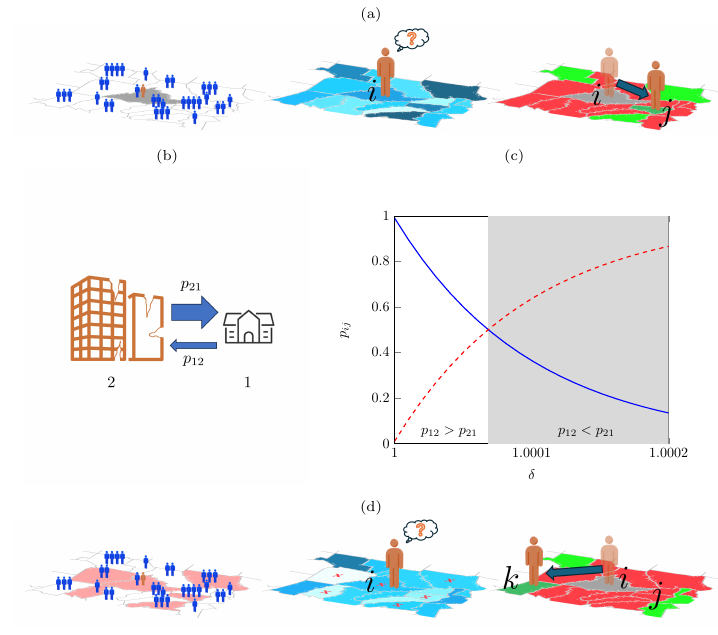}
		\caption{Failure of the standard radiation model to capture the effect of inequalities, and proposed improvement, the modified radiation model, to address this limitation. (a) Human decision-making process in the standard radiation model. Opportunities in each location are related to its population (left). When a person (orange) decides to commute or migrate from a region $i$ (gray), they set a benefit threshold based on the minimum opportunity they would accept to commute or migrate to another location. This threshold depends on opportunities at the origin $i$. A person evaluates opportunities in all possible destinations (center: a darker color indicates better opportunities for the person at a location; the color of the origin reflects the benefit threshold). Comparing these opportunities with their benefit threshold (right), they determine the regions whose opportunities exceed their threshold. In this way, they differentiate acceptable destinations (dark and light green) from unacceptable ones (red). The person will then commute or migrate to the closest acceptable destination, in this case $j$ (dark green). (b) Illustration of the 2-city system: city $1$ has a smaller population but is not affected by a conflict, and city $2$ has a much larger but affected by a conflict. (c) Probabilities $p_{12}$ of migration from city $1$ to $2$ (blue) and $p_{21}$ of migration from city $2$ to $1$ (red, dashed) from the modified radiation model (using $n_1 = 10^4$ and $n_2 = 10^6$), for varying values of the parameter $\delta$, which quantifies the inequality between cities $1$ and $2$. $\delta = 1$ corresponds to the standard radiation model, while increasing values greater than $1$ indicate a larger penalization of city $2$ compared to city $1$. The shaded area corresponds to the region of values of $\delta$ for which we have a higher propensity to move from the larger city $2$ to the smaller city $1$ -- a condition unattainable with the standard radiation model. (d) Human decision-making process in the modified radiation model. Contrary to the standard radiation model, opportunities in each location vary according to whether the location is suffering from a conflict or not (left; light red regions are part of the conflict). When a person (orange) decides to commute or migrate from a region $i$ that is part of the conflict, their benefit threshold is decreased compared to the standard radiation model (center: the color of the origin reflects the benefit threshold). Likewise, opportunities in destinations that suffer from the conflict are reduced compared to the standard radiation model (center: locations that are part of the conflict are marked with a red cross; a darker color indicates better opportunities for the person at a location). Comparing opportunities with the benefit threshold (right), the person moves to the closest acceptable destination, in this case $k$ (dark green), among those where opportunities exceed their threshold (dark and light green). Locations that are part of the conflict are more likely to offer opportunities worse than the threshold (red), modifying the acceptable destination landscape: for example, $j$, the destination in the standard radiation model, now becomes unacceptable. If the origin is part of the conflict, destinations that would not be acceptable in the standard radiation model can become acceptable: in this example, the destination $k$ was unacceptable in the standard radiation model.}
	\label{fig:1}
\end{figure*}

     A strong competitor to gravity models is the \textit{radiation model}, originally proposed by Simini \textit{et al.} \cite{simini2012universal}. Drawing inspiration from the intervening opportunity model \cite{stouffer1940intervening,schneider1959gravity} and radiation-absorption models used in physics \cite{kittel2018introduction}, the radiation model assumes knowledge of the number $T_i$ of commuters or migrants from each source location $i$. The model computes the number of them moving to each destination $j$. The individual decision-making process in the standard radiation model is illustrated in Fig. \ref{fig:1}(a). Each individual who commutes or migrates has a benefit threshold for mobility, which quantifies the minimum \textquotedblleft{}opportunity\textquotedblright{} a location must offer to them for considering to move. The individual moves to the closest destination whose opportunity is higher than their benefit threshold. Benefit threshold and opportunities are computed based on the premise that larger population centers tend to provide more opportunities than smaller ones. Thus, on average, the benefit threshold of a person in a larger city of origin will be higher; likewise, the opportunity of a more populous destination will be larger. The benefit threshold and opportunity are computed from a distribution $p(z)$, by drawing a number of extractions equal to the population of the location and taking the maximum value. In this way, populous cities will be a more desirable choice for human mobility.

     This individual-level model can be translated into average fluxes between locations with simple mathematical steps.  The flux from location $i$ to location $j$ (separated from a distance $r_{ij}$) with population $m_i$ and $n_j$, respectively, is related to the the probability $p_{ij}^\mathrm{R}$ that a commuting or migrating individual from $i$ will stop at location $j$,
    \begin{subequations}
    	\label{eq:Simini}
        \begin{equation}
		\label{eq:Tij_Simini}
		T_{ij} = T_ip_{ij}^{\mathrm{R}}, 
	\end{equation}
    \begin{equation}
 \label{eq:pij_Simini}
     p_{ij}^\mathrm{R} = 
		\frac{m_i n_j}{\left(m_i+s_{ij}\right)\left(m_i+n_j+s_{ij}\right)}.
 \end{equation}
    \end{subequations}
Here, $s_{ij}$ is the population within a radius $r_{ij}$ from the source location $i$, excluding source and destination. The flux in the radiation model is independent of the form of the benefit distribution $p(z)$.  The radiation model solves many of the issues of gravity models: it better captures long-range interactions; it is analytically consistent; and it is almost parameter-free (apart from the fraction of mobile population) \cite{simini2012universal}.

	The radiation model neglects inequalities between locations associated with living conditions and job quality. We embrace a broad definition of \textquotedblleft{}inequalities\textquotedblright{} between locations, one that includes the presence of conflicts, natural hazards, and political persecution \cite{van2020push}, along with socioeconomic disparities (such as income or wealth inequality) \cite{parkins2010push,obokata2014empirical,urbanski2022comparing,bilal2021scaling}. All of these variables can affect migration patterns. For example, consider an urban system with only two cities (1 and 2), one much more populous than the other ($n_2\gg n_1$), see Fig. \ref{fig:1}(b). Imagine that city 2 is affected by a conflict that poses its residents at risk: wouldn't we expect people from city 2 to flee to city 1? \eqref{eq:Simini} would suggest the opposite, whereby  $p_{12}^\mathrm{R}= n_2/(n_1+n_2) \gg p_{21}^\mathrm{R}=n_1/(n_1+n_2)$. Under the classical radiation model, only the population matters: one needs to account for inequalities across cities to be able to predict a higher tendency to move from $2$ to $1$, see Fig. \ref{fig:1}(c). Unfortunately, we have witnessed a similar situation during the ongoing Russian invasion of Ukraine, where over six million Ukrainians flee the country and settled in other nations, such as in Czech Republic \cite{Ukraine_crisis}, whose largest city, Prague, has less than half of Kyiv's population.

    Factors such as conflicts, natural hazards, and socioeconomic inequalities cannot be accounted for in the standard radiation model, where the benefit distribution is the same for all locations. Including inequalities between locations in mobility models is becoming a dire problem, with the increase of people on the move due to wars \cite{o2018migration} and climate change \cite{ionesco2016atlas} and potential increases in disparities in urban economies \cite{nijman2020urban}. For example, climate change does not impact all cities equally within the same urban system: some areas may be more affected than others for their specific location (such as coastal regions facing floods \cite{cazenave2014sea}) and the economies they have historically pursued (such as agricultural economies facing droughts \cite{cook2018climate}). Similarly, different cities have reacted differently to technological changes in their economies: Detroit and other cities in the so-called Rust Belt have been severely affected by the decline of US manufacturing \cite{wilson2022decline}, while metropolitan areas such as San Francisco and New York have thrived through the information economy \cite{drennan2002information}.
	
	Here, we propose a new mobility model that is intrinsically aware of inequalities among different locations (Fig. \ref{fig:1}d). The model assumes that locations can be divided in two  or more classes,  offering different levels of living conditions and opportunities. The classes have the same shape of the benefit distribution, but the distribution of class $j$ is modulated by a scaling factor $\delta_j$. Locations are assigned to different classes according to specific features that reflect inequalities, for example, the extent of a conflict. Contrary to other mobility models accounting for variables other than populations \cite{alis2021generalized,davis2018universal}, we do not alter the final equation of the radiation model in \eqref{eq:Simini}, but modify the human behavior model from which it is derived. With this approach, inequalities are incorporated in a natural way. In contrast to previously proposed schemes, this model does not require to estimate fluxes or add many, non-interpretable parameters -- an advantage over micro-founded gravity models as well. For the case of a uniform benefit distribution, we find closed-form expressions for the fluxes between locations  for an arbitrary number of classes.
	We present results on  migrations in South Sudan, driven by conflicts and natural hazards. We also apply the model to study the effects of socioeconomic inequalities on commuting patterns in the United States (US), to show how the proposed model outperforms the standard radiation model in widely different real-world datasets.
	
\section*{Results}

\subsection{Mathematical model}

To simplify the interpretation of formulas, we assume that the $L$ locations can be partitioned into two classes, where one class offers worse standard of living conditions and opportunities to their residents than the other (Fig. \ref{fig:1}d). The generalization to $C$ classes is presented in the Supporting Information. Residents can commute or migrate between any pair of locations, requiring to book-keep four distinct mobility patterns. This is a key difference from the standard radiation model \cite{simini2012universal} where all locations present equivalent opportunities, so that only one type of mobility pattern shall be resolved.

Within this working assumption, we describe opportunities of locations in each class through different probability mass functions $p^{(1)}(z)$ and $p^{(2)}(z)$, $z$ being the benefit threshold of an individual, equivalent to absorbance in physics. Functions describing classes of locations offering inferior opportunities are more ``concentrated'' towards the origin to favor the occurrence of worse opportunities. We assume that the second function is a scaled version of the first probability mass function, modulated by a parameter $\delta$: $p^{(2)}(z)=\delta p^{(1)}(z\delta)$ \footnote{This implies that the respective cumulative mass functions satisfy $P^{(2)}(z)=P^{(1)}(z\delta)$.}. The case $\delta=1$ corresponds to absence of inequalities, while $\delta>1$ indicates that the second class of locations offers worse opportunities than the first one. The parameter $\delta$ and the class of each location are assigned based on features that quantify inequalities between locations. While other choices of benefit distributions could be applied to penalize certain locations, the proposed approach allows the formulation of closed-form solutions, a critical aspect for applications in large datasets.
We introduce the following notation: $s_{ij}^{(k)}$ is the population of all locations of class  $k=1$ or $k=2$ (except of $i$ and $j$) within a circle of radius $r_{ij}$ centered at $i$; $s_{ij}$ is the population of all locations of any class (except of $i$ and $j$) within a circle of radius $r_{ij}$ centered at $i$  (such that $s_{ij}=s_{ij}^{(1)}+s_{ij}^{(2)}$); and $c_i$ represents the class of the $i$-th location.

Our micro-scale, individual behavioral model is analogous to the absorption-emission model of Simini \textit{et al.} \cite{simini2012universal}, consisting of two steps. First, we associate with each person $X$ who decides to commute or migrate  a given location $i$ a number $z_X$, representing their benefit threshold, computed as the maximum over $m_i$ extractions of the corresponding probability mass function $p^{(c_i)}(z)$. 
Second, the surrounding locations have a certain probability to absorb them, according to their size and class. For a generic location $j$, we compute its ``opportunity,'' analogously to the benefit threshold, by taking $n_j$ extractions from the corresponding probability mass function  $p^{(c_j)}(z)$. The person decides to move to the closest location whose opportunity is greater than its benefit threshold. In this modified model, the benefit distribution is not equal among locations, but also depends on the class of the location (that is, whether the location is disadvantaged with respect to others) and the extent $\delta$ to which inequality plays a role.

An individual decides to commute or migrate from location $i$ to location $j$ ($i\neq j$) when $j$ is the closest location to $i$ that has an opportunity larger than the benefit threshold of that individual. Thus, the probability of one person moving from location $i$ to location $j$ is
\begin{equation}\label{eq:Pea}
	\begin{split}
		&  P\left(1|m_i,n_j,s_{ij}^{(1)},s_{ij}^{(2)}\right)  \\&=
			\int_{0}^{\infty}\mathrm{d}z\, P_{m_i}(z)P_{s^{(1)}_{ij}}(<z)P_{s^{(2)}_{ij}}(<z)P_{n_{j}}(>z).
	\end{split}
\end{equation}
Each term in the integral can be easily interpreted. $P_{m_i}(z)$ is the probability that the maximum value extracted from the probability mass function corresponding to the origin location  ($p^{(c_i)}(z)$) after $m_i$ trials is equal to $z$, that is, that a person is assigned a benefit threshold $z$. Following Simini \textit{et al.}, such a quantity is computed as \footnote{This expression is easily derived by using the cumulative mass function $P_{m_i}(<z)$, which is equal to  $\left(p^{(c_i)}(<z)\right)^{m_i}$ due to the independence of the draws. By recalling that the probability mass function is equal to the derivative of its cumulative mass function with respect to its argument, the claim follows.}
\begin{equation}\label{eq:Pmi}
	 P_{m_i}(z)=
		m_ip^{(c_i)}(<z)^{m_i}p^{(c_i)}(z),
\end{equation}
 where $p^{(c_i)}(<z)$ represents the cumulative mass function of $p^{(c_i)}(z)$ (as all other draws must be below $z$ for $z$ to be the maximum). Likewise,  $P_{s^{(k)}_{ij}}(<z)$ ($k=1$ or $k=2$) is the probability that $s^{(k)}_{ij}$ numbers extracted from $p^{(k)}(z)$ are less than $z$: $p^{(k)}(<z)^{s_{ij}^{(k)}}$. The terms $P_{s^{(1)}_{ij}}(<z)$ and $P_{s^{(2)}_{ij}}(<z)$ in the integral require that no location of either class between $i$ and $j$ offers an opportunity above the benefit threshold $z$, so that the individual would pick that location as their destination. Finally, $P_{n_{j}}(>z)$ is the probability that among $n_j$ numbers extracted from the probability mass function corresponding to the destination location  ($p^{(c_j)}(z)$) at least one is greater than $z$,
\begin{equation}\label{eq:Pnj}
	 P_{n_j}(>z)= 1-p^{(c_j)}(<z)^{n_j}.
\end{equation}
This final term requires the opportunity of location $j$ to be higher than the benefit threshold $z$ of the individual, so that the individual would pick $j$ as their destination. We acknowledge that this form of lexicographic preference for the destination does not always represent human decision-making, but it offers a simple first-order approximation for the mathematical description of fluxes \cite{simini2012universal,barbosa2018human}. The integral sums the contributions for any possible benefit threshold extracted.

For each $i=1,\dots,L$, we set $p_{ii}=0$ (as a person who commutes or migrate will not stay in the same location) and scale the probabilities in \eqref{eq:Pea} such that $\sum_{j=1}^L p_{ij}=1$. The flux $T_{ij}$ from location $i$ to $j$ is computed by multiplying $p_{ij}$ by the known number of individuals $T_i$ from location $i$ who commute or migrate.

 \subsection{Closed-form results for two classes}  
 
 For the case of two classes \textquotedblleft{}1\textquotedblright{} and \textquotedblleft{}2\textquotedblright{}, there are four possible instances of \eqref{eq:Pea} ($1\rightarrow1$, $1\rightarrow2$, $2\rightarrow1$, and $2\rightarrow 2$). To perform the computations, we choose uniform distributions, which allow the formulation of a closed-form solution and represent the simplest ansatz for a benefit distribution, where benefits of commuting and migrating are uniformly distributed among the population. To limit the number of free parameters, we set the first class $p^{(1)}(z)=\mathrm{Rect}_1(z)$ to be the reference  and the second class as $p^{(2)}(z)=\delta\mathrm{Rect}_{\frac{1}{\delta}}(z)$, where $\mathrm{Rect}_a(z)$ is equal to 1 in $[0,a]$ and zero elsewhere with $a>0$. From simple algebra, we obtain

	\begin{subequations}\label{eq:pij}
		\begin{equation}
			p_{ij}^{(2\rightarrow 1)}=p_{ij}^{\mathrm{R}}\delta ^{-s^{(1)}_{ij}}\left[1+\frac{m_i+s_{ij}}{n_j}(1-\delta^{-n_j})\right],
		\end{equation}
		\begin{equation}
			p_{ij}^{(2\rightarrow 2)}=p_{ij}^{\mathrm{R}} \delta ^{-s^{(1)}_{ij}},
		\end{equation}
		\begin{equation}
			p_{ij}^{(1\rightarrow 2)}=p_{ij}^{\mathrm{R}} \delta ^{-m_i-s^{(1)}_{ij}},
		\end{equation}
		\begin{equation}
			\begin{split}
				& p_{ij}^{(1\rightarrow 1)}=  p_{ij}^{\mathrm{R}}\Bigg[\frac{s^{(2)}_{ij}(m_i+s_{ij}) \delta ^{-m_i-n_j-s^{(1)}_{ij}}}{n_j \left(m_i+n_j+s^{(1)}_{ij}\right) }\\&+ \frac{(m_i+s_{ij})(m_i+n_j+s_{ij})}{\left(m_i+s^{(1)}_{ij}\right) \left(m_i+n_j+s^{(1)}_{ij}\right)}\\&-\frac{ s^{(2)}_{ij}(m_i+n_j+s_{ij}) \delta ^{-m_i-s^{(1)}_{ij}}}{n_j\left(m_i+s^{(1)}_{ij}\right)}\Bigg].
			\end{split}
		\end{equation}
	\end{subequations}
	 These expressions are valid for any value of $\delta$. If $\delta=1$, the classes are equivalent and all expressions in \eqref{eq:pij} reduce to the standard radiation model \cite{simini2012universal}.  We have that $\delta>1$ ($\delta<1$) indicates that locations of class \textquotedblleft{}2\textquotedblright{} offer worse (better)  living conditions and opportunities than locations of class \textquotedblleft{}1\textquotedblright{}. In the Supporting Information, we compare fluxes of standard and modified radiation models for a small network and present results for the limit case $\delta\rightarrow \infty$. 
			
			 The expressions can be simplified for $\delta$ close to one through linearization, yielding $p_{ij}^{(k\rightarrow l)} = p_{ij}^{\mathrm{R}}\left[1+\rho_{ij}^{(k l)}(\delta-1)\right]$, with $\rho_{ij}^{21} = m_i+s_{ij}^{(2)}$, $\rho_{ij}^{22} = -s_{ij}^{(1)}$, $\rho_{ij}^{12} = -m_i-s_{ij}^{(1)}$, and $\rho_{ij}^{11} = s_{ij}^{(2)}$. The model and closed-form results can be extended to an arbitrary number of classes, where each class $k=1,\dots,C$ is assigned a parameter $\delta_k$ such that $p^{(k)}(z) = \delta_k p(z\delta_k)$, being $p(z)$ a reference benefit distribution, see Supporting Information. The linearized model also allows us to provide indications about the choice of the value of $\delta_k$, see Supporting Information.

		 \subsection{Analysis of South Sudan migration data}
   
\begin{figure*}[htbp]
			\centering
			\includegraphics[width=0.75\textwidth]{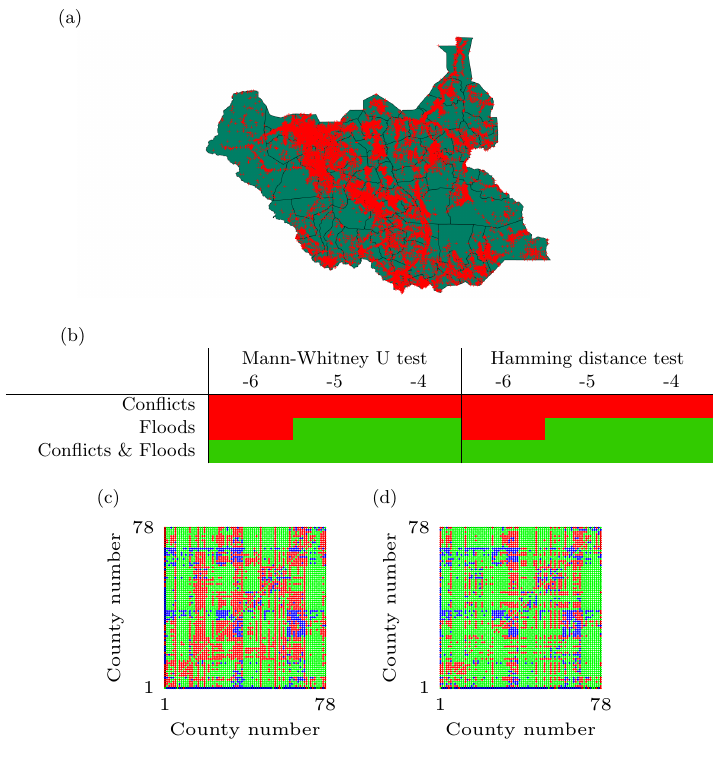}
			\caption{Results for South Sudan. (a) Main settlements areas in South Sudan in 2022, highlighted in red, based on data from Ref.~\cite{ocha_hdx}. (b) Results of statistical tests for the comparison between the standard and nonlinear modified radiation model with two classes for the South Sudan dataset, for each combination of variable of interest and value of $n$ in $\delta = 1+10^{n}$. Green results indicate significance in statistical tests and non-parametric comparisons ($\alpha = 0.05$). See Supporting Information for numerical results. (c-d) Instances where (c) the standard radiation model and (d) the modified radiation model accounting for both conflicts and floods ($\delta = 1+10^{-4}$) overestimate (red), correctly estimate (green), and underestimate (blue) the real migration flows recorded for South Sudan.}
			\label{fig:2}
		\end{figure*}

   We study 2020-2021 migrations in South Sudan (Fig. \ref{fig:2}a), a country plagued by civil war and environmental disasters such as floods, causing 2 million people to migrate internally over the past few years and 2.3 million refugees in other countries \cite{unhcr_South_Sudan}. This is an original dataset that we have collected and studied for the first time in the present work. Raw data from public sources have been polished, pre-processed, and made available to the scientific community in a dedicated online repository \cite{Online_Repository}.
		
		 As we anticipate large inequalities between counties, we utilize the nonlinear version of our model with two classes. The classes are selected depending on casualties in conflicts, scaled over the population, and intensity of floods in the counties (see Materials and Methods). Specifically, we explore three different possibilities to assign counties to the penalized class ($\delta>1$). Under the first option, penalized counties are the $15\%$ with higher casualties per capita. Under the second option, the penalized class includes the $15\%$ of counties that are mostly affected by floods. Under the third option, the penalized class is the union of the sets of penalized counties in the first two options. The choice of $15\%$ as the threshold for the most affected counties is a trade-off between two contrasting requirements: i) enough counties should be affected to achieve statistical differences; and ii) only a handful of counties are likely to be affected so severely as to obtain considerable changes in migration patterns. In the Supporting Information, we present a sensitivity analysis of the results with respect to this penalization threshold, highlighting the robustness of results to changes in the selection of this parameter. We repeat the analysis for different values of $\delta$ ($1+10^{n}$, $n = -6,-5,-4$) for the penalized class. We compare model predictions with those from the standard radiation model. For both the models, we assume that $T_i$ is known a priori, such that we only redistribute the overall flux from $i$ across all possible destinations, and we round fluxes to the closest integer.

		 For each estimated matrix of fluxes, we compute the error distribution by taking the absolute value of the difference between the estimated and real matrix of fluxes. We utilize two tests to assess whether the modified radiation model outperforms the standard radiation model. We employ the Mann-Whitney U test (MW) \cite{corder2014nonparametric} to test the null hypothesis that the absolute errors' distribution are statistically the same, with the alternative hypothesis being that the absolute errors of the modified radiation model are statistically lower than those of the standard radiation model. To provide an indication of whether the structure of the data generated by the models is similar to that of the actual fluxes, we use the Hamming distance (that is, the number of non-zero elements). This choice is motivated by the fact that most fluxes are zero, as most commuting and migration patterns are limited to neighboring locations. Specifically, we compute the Hamming distance for the error distributions of both modified and standard radiation models and statistically compare them (see Materials and Methods). We regard the MW and Hamming distance tests to be better indicators of the quality of the model than other indicators, such as least squares, root mean square, or the coefficient of determination, as they do not assume any relationship for the data (in particular, they are apt for nonlinear relationships) and better capture whether the results capture the overall structure of the mobility patterns. To exclude the possibility that improved model predictions are related to more parameters in the modified model, we employ a non-parametric test that penalizes the additional parameters in the model (see Materials and Methods). Specifically, we check whether a random assignment of the counties into two classes yields different predictions than assignment based on conflicts and floods. To further explore the role of the additional parameters in our modified model, we  replicate a statistical test with the Akaike information criterion (see Supporting Information). Only when statistics are significant and pass the non-parametric test (so that considered inequality variables actually provide information about mobility patterns), we consider the modified radiation model to outperform the standard one. For all tests, we use a significance level $\alpha=0.05$.

		 In Fig. \ref{fig:2}(b), we show the results of the comparisons of the modified radiation model against the standard one. We observe that conflicts, individually, cannot explain the recorded migration patterns. Only when considering floods or combining conflicts and floods we obtain significant results for both statistical tests and non-parametric comparisons. These results point at a critical role of climate and the environment on mobility. Further, we fail to identify significant results for individual variables for small values of $\delta$ for the penalized class, indicating that large penalties for locations affected by conflicts and floods are necessary.
		
		 Figure \ref{fig:2}(c-d) displays the comparison between the errors made by the standard radiation model and those of the modified radiation model considering both conflicts and floods, with $\delta = 1+10^{-4}$, against real flows. Most of the fluxes among counties are small (zero or a few individuals, especially at larger distances), such that both models are able to correctly estimate a good fraction of the fluxes. We find that the proposed model considerably increases the number of correctly estimated fluxes (in this case, $4,050$ against $3,366$ from the standard model), especially by reducing fluxes that are overestimated. In the Supporting Information, we present a rank comparison between fluxes of standard and modified radiation models against real fluxes.

		\subsection{Analysis of US commuting data} 
  
  As a second example of application of our model, we study commuting fluxes between US counties from 2011 to 2015. We examine a set of variables that are proxies of factors that reflect inequalities \cite{simpson2022demographic}, namely: i) Gini index (GI); ii) poverty ratio (PR); iii) ratio between median rent and median household income (RAT); and iv) unemployment rate (UR) (see Materials and Methods). For all variables, we consider averages over the five years.

		 As we expect smaller differences from the fluxes of the radiation model (due to the smaller effect of socioeconomic variables on migrations compared to conflicts and natural hazards) and larger variability in the variables of interest (due to the large number of counties) compared to the South Sudan dataset, we utilize the linearized model with an arbitrary number of classes. A higher value of each of the  socioeconomic variables is a proxy of higher inequality and, possibly, of worse job opportunities.  Thus, we set the number of classes equal to the number of counties whose variable is above the $90\%$ percentile, plus one. We assign to all counties with a value of the variable below the $90\%$ percentile to the first class, with $\delta_1=1$. For all the remaining counties, we set a value $\delta_k$ that varies linearly with the distance from the $90\%$ percentile, between $1$ (assigned to the county with the variable equal to the $90\%$ percentile) and $1+\Delta$ (assigned to the county with the maximum value of the variable). In this case, the choice of the $90\%$ percentile as a threshold is guided by the consideration that socioeconomic inequalities are likely to disrupt commuting patterns only in extreme conditions (unlike conflicts and natural hazards).

		For both the standard and modified radiation models, we compute $T_i$ from the population $m_i$ of each county using a linear regression of real commuting data. We find that $7.895\%$ of the population commutes outside their country of residence ($T_i = 0.0785\, m_i$).  We assign to $\Delta$ all powers of $10$ between $10^{-8}$ and $10^{-6}$. We utilize the same tests as those in the South Sudan case to compare the standard and modified radiation models (see Materials and Methods).

\begin{figure*}[htbp]
		\centering
		\includegraphics[width=0.7\textwidth]{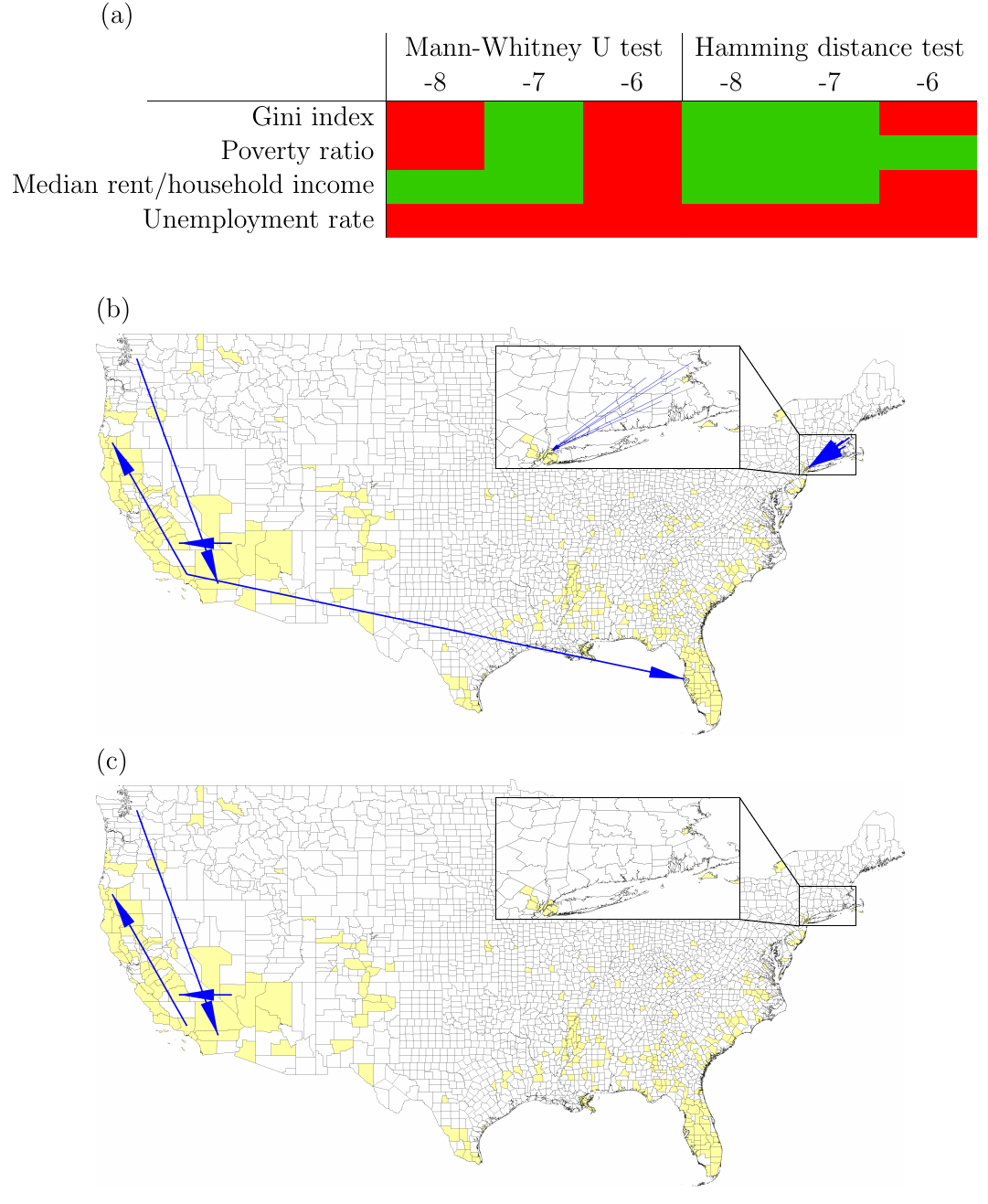}
		\caption{Results for the US. (a) Results of statistical tests for the comparison between the standard radiation model and linearized version of our model for the US dataset, for each combination of variable of interest and value of $n$ in $\Delta = 10^{n}$. Green results indicate significance in statistical tests and non-parametric comparisons ($\alpha = 0.05$). See Supporting Information for numerical results. (b-c)  Errors in fluxes for (b) the standard radiation model and (c) modified radiation model with RAT variable in US commuting patterns ($\Delta = 10^{-7}$). Only the ten fluxes that are changed the most in percentage by the modified model are shown. In (c), lack of arrows which were present in (b) indicate zero errors compared to real fluxes. All displayed fluxes are overestimated compared to real fluxes. Yellow counties correspond to the largest $10\%$ RAT values. The inset shows the area surrounding New York City}.
	\label{fig:3}
\end{figure*}

	 Figure \ref{fig:3}(a) shows the results of statistical tests comparing the linearized modified radiation models against the standard one. For moderate values of $\Delta$ ($10^{-7}$), modified models using GI, PR, and RAT outperform the standard radiation model, regardless of the statistical test. Interestingly, UR does not improve our predictions of commuting patterns, regardless of $\Delta$. It is tenable that differences in unemployment rates do not considerably affect commuting patterns, as in the short-term employment shocks are absorbed locally \cite{barkley2002estimating} and long-term variations often result in migrations \cite{bill2006examining}. Small ($10^{-8}$) and large values ($10^{-6}$) of $\Delta$  may hinder significant differences. For small $\Delta$-s,  effects may be too small to produce a significant difference, at least for MW tests. For large $\Delta$-s, the hypotheses of the linearized model are presumably not satisfied.

	In Fig. \ref{fig:3}(b-c), we show the errors in the ten fluxes that change the most in percentage values between the standard radiation model and modified radiation model using RAT and $\Delta = 10^{-7}$.  The standard radiation model overestimates several long-range commuting patterns from or to counties with high RAT, for example between Los Angeles and St. Petersburg in Florida and from Rhode Island/Massachusetts to counties forming New York City (Fig. \ref{fig:3}b). The modified radiation model correctly estimates most of these fluxes (as they are zero or comprising only a few individuals), removing overestimates in long-range commuting patterns that are absorbed by closer locations that are not affected by large RAT.

\section*{Discussion}

Mobility is a fundamental process in biological and non-biological systems, from morphogenesis of cells \cite{friedl2009collective} and animals' seasonal migrations \cite{fudickar2021animal}, to the motion of bacteria \cite{wadhams2004making} and colloids \cite{hong2007chemotaxis} through chemical gradients. Collective movements may support the emergence of higher-level, functional structures (such as embryos from cells \cite{aman2010cell}) or enable better exploitation of available resources (such as herds of animals migrating for food and reproduction \cite{alerstam2018ecology}). Human mobility is among the most complex of mobility processes \cite{havlin2012challenges}. Humans not only migrate long distances to settle where they can improve their living conditions  and escape conflicts and natural disasters, but also commute daily for work and leisure \cite{barbosa2018human}. The complexity of human mobility stems from the multitude of economic, social, and environmental factors underlying individual decision-making.

Current mathematical models of human mobility do not explicitly account for inequalities in living conditions and job opportunities. However, wars, natural hazards, political persecutions, and socioeconomic disparities have always been fundamental drivers of human mobility \cite{urbanski2022comparing}. With the increase in global conflicts \cite{o2018migration}, environmental crises from climate change \cite{ionesco2016atlas}, and potentially increasing inequalities in urban economies \cite{nijman2020urban}, there is a dire need for inequality-aware mobility models. To address this critical limitation, we propose a population-level model of human mobility, derived from first principles, that can account for inequalities that drastically affect human decision-making on mobility. We show that the proposed model can outperform the state-of-the-art radiation model in predicting migrations in South Sudan and commuting patterns in the US, modulating fluxes (especially long-range ones) from or to disadvantaged locations. We focus on statistical comparisons about the structure of the commuting patterns, without introducing additional hypotheses about relationships between variables.

The model contains two or more additional parameters that modulate the mobility patterns ($\delta_k$ and the threshold to distinguish disadvantaged locations from non-disadvantaged ones). We do not use optimization or fitting algorithms to select these parameters, as we are guided by order of magnitude arguments (see Supporting Information) as well as simple empirical considerations. For example, we select the threshold based on the need to include a sufficient number of disadvantaged locations, while retaining only locations for which the penalty effect is strong enough to see an effect, and a linear relationship between relevant inequality variables and penalties of different classes for $\delta_k$ as a simple ansatz. Yet, we acknowledge that the addition of parameters introduces additional complexity and arbitrariness. We addressed these concerns in several ways. First, we performed a non-parametric test to compare improvements in results when parameters are chosen according to relevant variable against instances where parameters are chosen randomly. If choosing parameters according to relevant variables do not significantly improve results against randomly assigning the parameters, we consider the model to not outperform the standard radiation model. Second, we replicated one of the statistical tests with an explicit penalization of the additional parameters by using the Akaike information criterion, obtaining analogous results (see Supporting Information). Finally, we showed that our results are robust to a change in the threshold with a robustness analysis (see Supporting Information).

The proposed model suffers from one main limitation: its results depend on the benefit distribution $p(z)$ considered. While a uniform distribution with finite support allows us to establish analytical results, other distributions may better capture benefit distributions, such as the Pareto distribution \cite{arnold2014pareto}. Pareto distributions comprise rare but very extreme events, thus possibly increasing some long-range fluxes that can only be absorbed by the largest cities with the best opportunities \cite{wang2024infrequent}. Thus, different distributions may capture inequalities at an individual level \cite{kirman1992whom,kirman2006heterogeneity}, on top of inequalities between locations. These and other non-rectangular $p(z)$ are adapt to numerical implementations of the modified radiation model. The modifications introduced in this manuscript can be also applied to more advanced radiation models \cite{simini2013human,yang2014limits}. Other fundamental hypotheses of the proposed model are the lexicographic preferences for the destination location and the form of the scaling of the benefit distribution for disadvantaged locations. While these choices may not always accurately represent human decision-making in mobility processes, we decided to retain them, together with uniform distributions, to guarantee the formulation of closed-form solutions for the fluxes. We consider closed-form solutions fundamental to reduce the computational complexity of the model, especially in view of the large datasets associated with mobility patterns (which scale quadratically with the number of locations).

A critical application of the proposed model is the study of inequalities associated with climate change, which will likely cause unprecedented environmental migrations in the next decades \cite{de2021modeling,wolde2023environmental}. This topic is the object of intense political debate \cite{Guardian_climate,Center_for_Global_Development} and is in dire need of scientifically backed evidence. The study of the South Sudan dataset points at a crucial role of climate inequalities on human migrations associated with exposure to floods. At the moment, the field lacks reliable datasets that track where people moves in the aftermath of extreme events (and if and how they return) \cite{wang2014quantifying,yabe2022mobile,yabe2024enhancing}. The South Sudan dataset, which we collated, is a first step to bridge this knowledge gap. We anticipate that new datasets, along with the proposed model, will help predict the effect of climate change on future human migrations. Such predictions are of critical importance for cities, the most sought-after destinations of migrants \cite{monras2023immigration,adamo2010environmental}. These analyses can equip cities' governance with adequate information tools to prepare for stresses on their infrastructural and service systems.

\section*{Materials and Methods}

\subsection*{Data of South Sudan migrations}

South Sudan gained independence from Sudan in July 2011. In December 2013, a civil war broke out, and the conflict officially ended in September 2018 with the signing of the ``Revitalized Agreement on the Resolution of the Conflict in South Sudan'' (R-ARCSS)~\cite{onapa2019south}. Five years of war forced nearly 4.2 million people to flee their homes in search of safety. Of these 4.2 million people, about a half remained in the country -- so called Internally Displaced People (IDPs), according to the UN\footnote{UN defines IDPs as ``persons or groups of persons who have been forced or obliged to flee or to leave their homes or places of habitual residence, in particular as a result of or in order to avoid the effects of armed conflict, situations of generalized violence, violations of human rights or natural or human-made disasters, and who have not crossed an internationally recognized state border''~\cite{guiding_principles}.}. In 2020, 1.6 million of these IDPs were still displaced, mainly because of communal clashes and other forms of violence against civilians. In 2021, such a figure further increased, to about 2 million people, partly due to floods that occurred in the country since 2019; for example, 835,000 people were affected by flooding between May and December 2021~\cite{OCHA_southsudan_2022}.

\subsubsection*{Migration data}
The International Organization for Migration (IOM) collects data about displaced people in South Sudan through the Displacement Tracking Matrix (DTM) system, encompassing mutiple datasets~\cite{dtm_datasets}. To study internal movements in the country, we used the ``Flow Monitoring'' dataset, which provides quantitative estimates of the flow of individuals through specific locations. In particular, the Flow Monitoring Registry (FMR) surveys people on the move during selected hours of observations at key transit points (Flow Monitoring Points -- FMPs), within South Sudan and on its borders. FMR datasets are organized by month starting from 2020, and for each observed displacement they include the following information:
\begin{itemize}
	\item Name and position of the FMP where the data was collected;
	\item Total number of individuals in the surveyed group, and number of those who declare South Sudanese nationality;
	\item Type of travel relative to South Sudan (incoming, outgoing, internal, or transit);
	\item Country of departure and (intended) destination, including level 1 and 2 administrative sub-areas;
	\item Camp of departure or destination, if any;
	\item Time spent by the group at the location of departure and intended time to be spent at the destination, used to distinguish medium/long term displacements from others;
	\item Whether or not the destination matches the group's habitual residence;
	\item Reason for displacement, differentiating between voluntary and forced ones; and
	\item Means of transport.
\end{itemize}
We used data from January 2020 to December 2021. Note that data in FMR mostly refer voluntary, short-term migrations, even though some forced migrations are still registered. Moreover, FMR data only capture the volume and characteristics of the flows transiting through the FMPs, and they do not provide a full or statistically representative picture of internal and cross-border movements in South Sudan.

\subsubsection*{Population data}
In the absence of a recent census, we used estimates of the South Sudan population annually as part of the Humanitarian Programme Cycle analysis carried out by the UN High Commissioner for Refugees. These estimates are developed and endorsed by the Common Operational Datasets for population statistics (COD-PS), and disseminated through the OCHA's Humanitarian Data Exchange website (HDX) \cite{ocha_hdx}. CODs are the reference datasets to support operations and decision-making in the initial response to a humanitarian emergency.  Only estimates of South Sudanese nationals in South Sudan are included in the COD-PS dataset, thereby excluding refugees and asylum seekers in and out of the country. In this work, population estimates in 2020 were linked to the administrative boundaries (AB) in the COD-AB dataset, also available at the HDX website \cite{ocha_hdx}.

In our work, we also utilized settlements data at the subnational level to obtain more accurate information about the population distribution in the country. These data are also available at the HDX website, and they comprise geographical coordinates of settlements with their administrative sub-areas from level 1 to 3.

\subsubsection*{Natural disasters and conflicts}
As we intended to model the impact of natural disasters and conflicts on internal migrations, we also considered two datasets on flooding and violence in South Sudan. With respect to flooding, for the year 2021, we employed the OCHA dataset on the number of people affected by floods at the county level. For 2020, data on flooding were not available from HDX, so an estimate was obtained from Figure 1 in \cite{OCHA_southsudan_2022} that visualizes the number of people affected by floods in each county with a resolution of 25,000 people. We chose the middle value of the county bin as an estimate of the number of affected people. 

With respect to conflicts, data were collected from the Armed Conflict Location \& Event Data Project (ACLED) \cite{acled}, which reports information about the type, agents, location, date, and other characteristics of political violence events around the world. In this work, we focused on the number of casualties in South Sudan due to violence against civilians at the level of the county with a resolution of one year. We aggregated values over 2020 and 2021 and scaled the number of casualties by the population, to obtain a relative probability of civilians being affected by the conflict.

\subsubsection*{Preprocessing of the data}

A county-level origin-destination (OD) matrix is built by using movements described in FMR data. The ones that are internal to South Sudan are extracted from the available datasets and only South Sudanese nationals who moved from one county to another are taken into account, filling the element of the matrix corresponding to their reported origin and intended destination. A single OD matrix is obtained by adding the displacements that occurred in all months of 2020 and 2021. In this case, $965$ non-zero flows are present.

Settlements data and COD-AB boundaries are used to compute the geographical coordinates of the centroid of each county. In particular, settlements within the boundaries of each county are extracted and their coordinates are averaged to compute the centroid of that county. 

\subsection*{Statistical test for South Sudan}

We utilize the MW statistical test to compare the results of the standard radiation model against the modified radiation model utilizing different variables of interest and values of $\Delta$. The null hypothesis of the test is that the absolute errors' distributions for standard and modified radiation models are statistically indistinguishable. On the other hand, the alternative hypothesis requires the absolute errors of the modified radiation model to be statistically lower than those of the standard radiation model.

We seek to assess whether the improvement in the MW test is only related to a higher number of parameters ($\delta$ and the percentile of the distribution of variables that separates the two classes of locations). To this end, we compare the results of the modified radiation model against an empirical distribution of the same statistics. To generate this empirical distribution, we generate fluxes via the modified radiation model with the same $\delta$, with classes assigned randomly while maintaining the same number of counties in each class. We utilize these fluxes to perform a MW statistical test against the standard radiation model. We repeat this procedure 1,000 times to generate a distribution of statistics from the test. Next, we perform a non-parametric test by comparing the statistics from the test with the modified model with relevant variables against the $5\%$ percentile of the empirical distribution.

We repeat the same procedure for the test on the Hamming distance on the error between each model and real fluxes. To assess whether the improvement in the Hamming distance in the modified radiation model is significant, we utilize a statistical test. We model the Hamming distance as the outcome of random extractions from a binomial distribution over $N^2$ trials (being $N$ the number of counties), with success probability estimated from the Hamming distance for the standard radiation model. The null hypothesis is that the Hamming distance observed in the modified radiation model is generated from the same binomial distribution of the standard one. The alternative hypothesis is that there are two different binomial distributions generating the data, where the success probability of the binomial distribution for the modified radiation model is lower than that of the standard one. We use as $p$-value for the test the probability that the Hamming distance is equal or smaller than that observed in the modified radiation model (that is, the cumulative binomial distribution), assuming the success probability to be equal to that estimated from the standard radiation model.

To penalize the additional parameters in the model, we compare Hamming distances from modified radiation models against an empirical distribution of Hamming distances. Similar to the previous test, we generate fluxes from the modified radiation model by assigning counties to the two classes randomly, while maintaining the same ratio of locations in each class. Next, we compute the Hamming distance for the error between these fluxes and the real fluxes. Repeating the procedure $1,000$ times, we generate the empirical distribution of Hamming distances, from which we extract the $5\%$ percentile, which we use to perform a non-parametric comparison.

\subsection*{Data of commuting patterns and socioeconomic variables for the United States}

Commuting fluxes between US counties from 2011 to 2015 are available from the American Community Survey \cite{ACS_commuting_data}. The variables utilized as proxies of socioeconomic inequalities, available from the Agency for Healthcare Research and Quality \cite{AHRQ_data}, are defined as follow:
\begin{itemize}
    \item Gini index (GI): Gini index of income inequality \cite{de2007income}. It is a measure of income inequality based on the Lorenz curve, which plots the cumulative percentage of the overall income earned by the bottom $x\%$ of the population. For a perfectly equally distributed income, the curve would be a straight, $45\degree$ line, called line of equality. In practice, the real curve always lays below the line of equality. The Gini index is the ratio between the area between the line of equality and the Lorenz curve, over the entire area below the line of equality. The index goes from a minimum of 0 (the entire income is concentrated in a single individual) to a maximum of 1 (the income is equally distributed among all individuals).
    \item Poverty ratio (PR): Percentage of population with income to poverty ratio under 0.50. The poverty line is defined by U.S. Census Bureau (\$27,479 for a family of four in 2021 \cite{Poverty_data}). Higher values indicate larger portions of the population below the poverty line.
    \item Ratio between the median rent and median household income (RAT): Ratio between the annualized median gross rent (dollars) and the median household income (dollars, inflation-adjusted to data file year). Higher values indicate that residents pay a larger amount of the household income in rent.
    \item Unemployment rate (UR): Unemployment rate per 100 population (ages 16 and over).
\end{itemize}

The plots of the counties over the 90$\%$ percentile for GI, PR, RAT, and UR are available in the Supporting Information.

\subsection*{Statistical tests for the United States}

We replicate the same statistical tests proposed for South Sudan in the case of US counties, with a few differences. First, in the MW tests, as the vast majority of commuting fluxes are zero, we remove from the distributions the values for which both the errors of the standard and modified radiation models are zero to emphasize the differences between the methods. Second, we limit random extractions in the non-parametric tests to 200, due to the large number of counties (generation of the random dataset employed about a week on a standard laptop).

\begin{acknowledgments}
	The authors acknowledge financial support from the National Science Foundation under grant No. CMMI-2332144, EF-2222418, and New York University’s Mega Seed Grant No. A21-0738.
\end{acknowledgments}

\section*{Supporting Information}

The Supporting Information contains: an explicative application of our model to a small graph; the limit behavior of the two-class model for extreme inequalities; the generalization of nonlinear and linearized models to an arbitrary number of classes; order of magnitude arguments that guide the selection of the penalty parameters $\delta_k$; numerical results for South Sudan, along with a rank comparison of fluxes, a sensitivity analysis on the penalization threshold and the replication of a statistical test with the Akaike Information Criterion to penalize for additional model parameters; plots of counties with the most extreme values of the variable of interests in the US; and numerical results for the US.

\subsection*{Data availability}

The dataset on South Sudan is available at \cite{Online_Repository}.

\subsection*{Author contributions}
Conceptualization: A.B., P.D.L., R.D., L.C., M.H., M.P.; Investigation: A.B., P.D.L., R.D., L.C., M.H., M.P.; Methodology: A.B., P.D.L., S.I., M.P.; Software: A.B., P.D.L., S.I., M.P.; Validation: A.B.; Formal Analysis: A.B., M.P.; Data Curation: A.B., P.D.L., S.I.; Project Administration: A.B.; Writing - Original Draft: A.B., P.D.L., M.P.; Writing - Review \& Editing: A.B., P.D.L., R.D., L.C., M.H., M.P.; Supervision: M.P.; Funding acquisition: L.C., M.P.

\subsection*{Conflict of interests statement}
The authors declare no competing interests.

\newpage

\renewcommand\theequation{S\arabic{equation}}
\setcounter{equation}{0}
\renewcommand{\thefigure}{S\arabic{figure}}
\setcounter{figure}{0}
\renewcommand\thetable{S-\Roman{table}}
\setcounter{table}{0}

\begin{widetext}
\begin{center}
	\textbf{\Large Supporting Information}	
\end{center}

\section{Comparison on a small graph} 

We illustrate how the proposed model differs from the standard radiation model on a small graph, composed of ten nodes with a random position in $[0,1]\times[0,1]$ and random population (Fig. \ref{fig:S1}). We arbitrary label the nodes of the graph and assign the first three of them to the \textquotedblleft{}2\textquotedblright{} class, to signify that they are experiencing detrimental effects from a variable (for example, they have higher unemployment rates). The rest of the nodes are assigned to the \textquotedblleft{}1\textquotedblright{} class. 

We run a series of simulations by varying the parameter $\delta$. Specifically, we consider the values $\delta=1$ (equivalent to the standard radiation model), $1.01$, and $1.05$, corresponding to increasing inequities between affected and unaffected locations, and we assume that $25\%$ of the population moves from each node. Population fluxes for different values of $\delta$ are shown in Fig. \ref{fig:S1}(b-d).

The standard radiation model (Fig. \ref{fig:S1}b) shows large fluxes between the closest communities, that is, the pairs $(2,6)$ and $(4,7)$, with a proportional bias toward the larger community of the pair, 6 and 7, respectively. All of the nodes display a sizeable flux toward 1, which is the largest and most central community.

Upon disrupting communities 1, 2, and 3, we discover substantial changes in mobility patterns. Already for $\delta=1.01$ (Fig. \ref{fig:S1}(c)) we find a decrease in all the fluxes from the rest of the nodes to these communities. We still register a consistent flux from 3 to 1, due to the relative isolation of 3 and the large size of 1. We find a considerable increase in the flux from 1 to the closest non-affected community, 5, and to 6, which is considerably larger than 5, although a bit further. In addition, migration fluxes from 2 to 6 become much larger, but there is still a considerable number of migrants from 6 to 2, despite the inequalities in 2. Fluxes between unaffected nodes are also affected. Most notably, we find larger fluxes from community 6 to 5, likely redirected away from 1 and 2, and between 6 and 7.

For $\delta=1.05$ (Fig. \ref{fig:S1}(d)), we find that almost all fluxes from unaffected communities to affected ones are almost negligible. Interestingly, there is still a significant flux from node 3 to 1. Despite the disruption in 1, most of the migrants from 3 still decide to move to 1, as there is no closer unaffected location (see Section \ref{sec:limit_model_delta_infinity}). Unexpectedly, we find that some fluxes from affected communities to unaffected ones increase, while others decrease. In particular, while for smaller values of $\delta$ we record an increase of the fluxes from 1 to 5 and from 1 to 6, at larger values we find that the flux from 1 to 5 still increases, while that from 1 to 6 decreases. In fact, for large $\delta$-s, the population in a disadvantaged location tends to migrate only to the closest, unaffected location, regardless of the population, as they seek to get away in the fastest possible way from the affected location (see Section \ref{sec:limit_model_delta_infinity}). Fluxes from unaffected locations 6 and 7 shift from the affected nodes 1 and 2 to the closest unaffected ones, 4 and 5, and long-range migration patterns between 6 and 7 are further strengthened.

\begin{figure*}[htb]
	\centering
	\includegraphics[width=.75\textwidth]{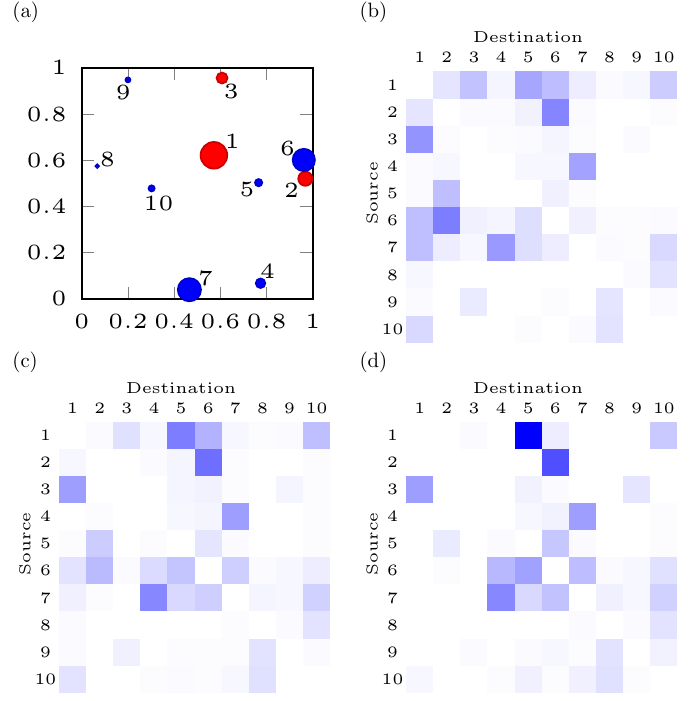}
	\caption{Comparison of the proposed model and standard radiation model on a small graph. (a) Spatial location and labeling of the nodes, whose size is proportional to the population (red: \textquotedblleft{}2\textquotedblright{}, blue: \textquotedblleft{}1\textquotedblright{}). (b-d) Population fluxes between nodes for $\delta=1$ (equivalent to the standard radiation model), $1.01$, and $1.05$, respectively; larger population fluxes correspond to darker shades, with an equal color scale among the three panels.}
	\label{fig:S1}
\end{figure*}

\section{Limit of the model for $\delta\rightarrow\infty$}
\label{sec:limit_model_delta_infinity}

An interesting case for our model is that in which $\delta\rightarrow\infty$, which corresponds to a case in which inequalities are so stark that affected locations basically offer no opportunity and do not attract any migrant or commuter. In this case, for the nonlinear model with two classes and arbitrarily large $\delta$, we find
\begin{subequations}
	\begin{equation}
		p_{ij}^{(2\rightarrow 1)} = \begin{cases}
			0 \,\,\,\text{if } s_{ij}^{(1)}>0, \\
			\frac{m_i}{m_i+s_{ij}^{(2)}} \,\,\,\text{if } s_{ij}^{(1)}=0,
		\end{cases},
	\end{equation}
	\begin{equation}
		p_{ij}^{(2\rightarrow 2)} = \begin{cases}
			0 \,\,\,\text{if } s_{ij}^{(1)}>0, \\
			\frac{m_in_j}{\left(m_i+s_{ij}^{(2)}\right)\left(m_i+n_j+s_{ij}^{(2)}\right)} \,\,\,\text{if } s_{ij}^{(1)}=0,
		\end{cases},
	\end{equation}
	\begin{equation}
		p_{ij}^{(1\rightarrow 2)} = 0,
	\end{equation}
	\begin{equation}
		p_{ij}^{(1\rightarrow 1)} = \frac{m_i n_j}{\left(m_i+s_{ij}^{(1)}\right)\left(m_i+n_j+s_{ij}^{(1)}\right)}.
	\end{equation}
\end{subequations}

We find that the unaffected locations follow the standard radiation model as if there was no affected location. There is no flux from unaffected to affected locations. In affected locations, the entire population is emitted. We find that an affected location only have non-zero fluxes to locations within the radius of the closest unaffected location. Within that radius, affected locations exchange fluxes between themselves following the standard radiation model. Thus, small, isolated, disadvantaged communities surrounded by other affected locations would still migrate to communities with larger populations, even if disadvantaged as well. The rest of the population is absorbed by the closest unaffected location.

\section{Nonlinear model for an arbitrary number of classes}

We present the derivation of our model for an arbitrary number of classes. To find closed-form expressions, we use uniform distributions, such that $p^{(c_i)}(z) = \delta_{c_i} \mathrm{Rect}_{\delta_{c_i}^{-1}}(z)$ and the corresponding cumulative mass function is
\begin{equation}
	P^{(c_i)}(<z) = \delta_{c_i}\mathrm{Ramp}_{\delta_{c_i}^{-1}}(z),
\end{equation}
with
\begin{equation}
	\mathrm{Ramp}_{\delta_{k}^{-1}}(z) = \begin{cases}
		z, \hspace*{16pt} 0\leq z < \delta_{k}^{-1} \\
		\delta_{k}^{-1}, \hspace*{16pt} z\geq \delta_{k}^{-1}
	\end{cases}.
\end{equation}
We establish

	\begin{equation}
		\begin{split}
			& P\left(1|m_i,n_j,s_{ij}^{(1)},\dots,s_{ij}^{(C)}\right)  =
			\int_{0}^{\infty}\mathrm{d}z P_{m_i}(z)\prod_{k=1}^C  \left[P_{s^{(k)}_{ij}}(<z)\right]P_{n_{j}}(>z) \\& = m_i \delta_{c_i}^{m_i} \prod_{k=1}^C \delta_k^{s_{ij}^{(k)}} \int_0^\infty \left[\mathrm{Ramp}_{\delta_{c_i}^{-1}}(z)\right]^{m_i} \mathrm{Rect}_{\delta_{c_i}^{-1}}(z) \left[\mathrm{Ramp}_{\delta_{k}^{-1}(z)}\right]^{s_{ij}^{(k)}}\,\mathrm{d}z \\
			& - m_i \delta_{c_i}^{m_i}\delta_{c_j}^{n_j} \prod_{k=1}^C \delta_k^{s_{ij}^{(k)}} \int_0^\infty \left[\mathrm{Ramp}_{\delta_{c_i}^{-1}}(z)\right]^{m_i} \mathrm{Rect}_{\delta_{c_i}^{-1}}(z) \left[\mathrm{Ramp}_{\delta_{k}^{-1}(z)}\right]^{s_{ij}^{(k)}}\left[\mathrm{Ramp}_{\delta_{c_j}^{-1}(z)}\right]^{n_j}\,\mathrm{d}z.
		\end{split}
	\end{equation}

Without lack of generality, let us define $\delta_1<\delta_2<\dots<\delta_C$. We first focus on the first term. By carrying out the integration, we write
	\begin{equation}
		\begin{split}
			\label{eq:first_term_nonlinear}
			& m_i \delta_{c_i}^{m_i} \prod_{k=1}^C \delta_k^{s_{ij}^{(k)}} \int_0^\infty \left[\mathrm{Ramp}_{\delta_{c_i}^{-1}}(z)\right]^{m_i} \mathrm{Rect}_{\delta_{c_i}^{-1}}(z) \left[\mathrm{Ramp}_{\delta_{c_k}^{-1}(z)}\right]^{s_{ij}^{(k)}}\,\mathrm{d}z \\
			& = m_i \delta_{c_i}^{m_i} \prod_{k=1}^C \delta_k^{s_{ij}^{(k)}} \left[\dfrac{\delta_C^{-(m_i+s_{ij})}}{m_i+s_{ij}}+\sum_{r=c_i}^{C-1}\dfrac{1}{\prod_{s=r+1}^C\delta_s^{s_{ij}^{(s)}}}\dfrac{\delta_r^{-\left(m_i+\sum_{t=1}^r s_{ij}^{(t)}\right)}-\delta_{r+1}^{-\left(m_i+\sum_{t=1}^r s_{ij}^{(t)}\right)}}{m_i+\sum_{t=1}^r s_{ij}^{(t)}}\right].
		\end{split}
	\end{equation}

Next, we focus on the second term. Assuming $c_i<c_j$, the integration provides
	\begin{equation}
		\begin{split}
			\label{eq:second_term_nonlinear}
			& m_i \delta_{c_i}^{m_i}\delta_{c_j}^{n_j} \prod_{k=1}^C \delta_k^{s_{ij}^{(k)}} \int_0^\infty \left[\mathrm{Ramp}_{\delta_{c_i}^{-1}}(z)\right]^{m_i} \mathrm{Rect}_{\delta_{c_i}^{-1}}(z) \left[\mathrm{Ramp}_{\delta_{k}^{-1}(z)}\right]^{s_{ij}^{(k)}}\left[\mathrm{Ramp}_{\delta_{c_j}^{-1}(z)}\right]^{n_j}\,\mathrm{d}z \\ & = m_i \delta_{c_i}^{m_i}\delta_{c_j}^{n_j}\prod_{k=1}^C \delta_k^{s_{ij}^{(k)}}\Bigg[\dfrac{\delta_K^{-(m_i+n_j+s_{ij})}}{m_i+n_j+s_{ij}} + \sum_{r=c_j}^{C-1}\dfrac{1}{\prod_{s=r+1}^C \delta_s^{s_{ij}^{(s)}}}\dfrac{\delta_r^{-\left(m_i+n_j+\sum_{t=1}^r s_{ij}^{(t)}\right)}-\delta_{r+1}^{-\left(m_i+n_j+\sum_{t=1}^r s_{ij}^{(t)}\right)}}{m_i+n_j+\sum_{t=1}^r s_{ij}^{(t)}}\\& + \sum_{r=c_i}^{c_j-1}\dfrac{1}{\prod_{s=r+1}^C \delta_s^{s_{ij}^{(s)}}}\dfrac{\delta_r^{-\left(m_i+\sum_{t=1}^r s_{ij}^{(t)}\right)}-\delta_{r+1}^{-\left(m_i+\sum_{t=1}^r s_{ij}^{(t)}\right)}}{m_i+\sum_{t=1}^r s_{ij}^{(t)}}\Bigg].
		\end{split}
	\end{equation}
For $c_i\geq c_j$, the third term in parentheses is zero, while the summation in the second term starts from $c_i$. 

Merging the results in \eqref{eq:first_term_nonlinear} and \eqref{eq:second_term}, we find
	\begin{equation}
		\begin{split}
			\label{eq:nonlinear_arbitrary_delta}
			& p_{ij}^{c_i\rightarrow c_j} = m_i \delta_{c_i}^{m_i} \prod_{k=1}^C \delta_k^{s_{ij}^{(k)}} \Bigg[ \dfrac{\delta_C^{-(m_i+s_{ij})}}{m_i+s_{ij}} - \delta_{c_j}^{n_j}\dfrac{\delta_K^{-(m_i+n_j+s_{ij})}}{m_i+n_j+s_{ij}} \\ &+ \sum_{r=c_i}^{C-1}\dfrac{1}{\prod_{s=r+1}^C\delta_s^{s_{ij}^{(s)}}}\dfrac{\delta_r^{-\left(m_i+\sum_{t=1}^r s_{ij}^{(t)}\right)}-\delta_{r+1}^{-\left(m_i+\sum_{t=1}^r s_{ij}^{(t)}\right)}}{m_i+\sum_{t=1}^r s_{ij}^{(t)}} \\ & -\delta_{c_j}^{n_j} \Bigg(\sum_{r=c_j}^{C-1}\dfrac{1}{\prod_{s=r+1}^C \delta_s^{s_{ij}^{(s)}}}\dfrac{\delta_r^{-\left(m_i+n_j+\sum_{t=1}^r s_{ij}^{(t)}\right)}-\delta_{r+1}^{-\left(m_i+n_j+\sum_{t=1}^r s_{ij}^{(t)}\right)}}{m_i+n_j+\sum_{t=1}^r s_{ij}^{(t)}}\\& + \sum_{r=c_i}^{c_j-1}\dfrac{1}{\prod_{s=r+1}^C \delta_s^{s_{ij}^{(s)}}}\dfrac{\delta_r^{-\left(m_i+\sum_{t=1}^r s_{ij}^{(t)}\right)}-\delta_{r+1}^{-\left(m_i+\sum_{t=1}^r s_{ij}^{(t)}\right)}}{m_i+\sum_{t=1}^r s_{ij}^{(t)}}\Bigg)\Bigg].
		\end{split}
	\end{equation}

\subsection{Linearized model for an arbitrary number of classes}

Due to the complexity of the formula in \eqref{eq:nonlinear_arbitrary_delta}, we put forward a linearized version that is fully transparent and interpretable. We linearize the first term in \eqref{eq:first_term_nonlinear} for $\delta_k\rightarrow 1$, $k=1,\dots,C$. The zero-th order term is $m_i/(m_i+s_{ij})$. The first-order terms in $\delta_k$ depend on the relation between $\delta_k$ and $\delta_{c_i}$. We have that, for $\delta_k\neq \delta_{c_i}$, the first-order terms are in the form
\begin{equation}
	\dfrac{m_i}{m_i+s_{ij}}s_{ij}^{(k)}(\delta_k-1).
\end{equation}
If instead $\delta_k= \delta_{c_i}$, we find the first-order term is
\begin{equation}
	\dfrac{m_i}{m_i+s_{ij}}(s_{ij}^{(c_i)}-s_{ij})(\delta_{c_i}-1).
\end{equation}
We can rewrite the linearization as 
\begin{equation}
	\begin{split}
		\label{eq:first_term}
		& m_i \delta_{c_i}^{m_i} \prod_{k=1}^C \delta_k^{s_{ij}^{(k)}} \int_0^\infty \left[\mathrm{Ramp}_{\delta_{c_i}^{-1}}(z)\right]^{m_i} \mathrm{Rect}_{\delta_{c_i}^{-1}}(z) \left[\mathrm{Ramp}_{\delta_{c_k}^{-1}(z)}\right]^{s_{ij}^{(k)}}\,\mathrm{d}z \\
		& \approx \dfrac{m_i}{m_i+s_{ij}} \left[1+\sum_{k=1}^C s_{ij}^{(k)}(\delta_k-1)-s_{ij}(\delta_{c_i}-1)\right].
	\end{split}
\end{equation}

We then linearize the second term in \eqref{eq:second_term_nonlinear}. The zero-th order term of the sum is $m_i/(m_i+n_j+s_{ij})$. Similar to the previous case, the first order terms depend on the relation between $\delta_k$, $\delta_{c_i}$, and $\delta_{c_j}$. For $\delta_k\neq \delta_{c_i}$ and $\delta_k\neq \delta_{c_j}$, the first-order terms are in the form
\begin{equation}
	\dfrac{m_i}{m_i+n_j+s_{ij}}s_{ij}^{(k)} (\delta_k-1).
\end{equation}
For $\delta_k=\delta_{c_i}$,
\begin{equation}
	\dfrac{m_i}{m_i+n_j+s_{ij}}(s_{ij}^{(c_i)}-s_{ij}-n_j)(\delta_{c_i}-1).
\end{equation}
For $\delta_k=\delta_{c_j}$,
\begin{equation}
	\dfrac{m_i}{m_i+n_j+s_{ij}}(s_{ij}^{(c_j)}+n_j)(\delta_{c_j}-1).
\end{equation}
Then, we write the linearization as
\begin{equation}
	\label{eq:second_term}
	\begin{split}
		& m_i \delta_{c_i}^{m_i}\delta_{c_j}^{n_j} \prod_{k=1}^C \delta_k^{s_{ij}^{(k)}} \int_0^\infty \left[\mathrm{Ramp}_{\delta_{c_i}^{-1}}(z)\right]^{m_i} \mathrm{Rect}_{\delta_{c_i}^{-1}}(z) \left[\mathrm{Ramp}_{\delta_{k}^{-1}(z)}\right]^{s_{ij}^{(k)}}\left[\mathrm{Ramp}_{\delta_{c_j}^{-1}(z)}\right]^{n_j}\,\mathrm{d}z \\ & \approx \dfrac{m_i}{m_i+n_j+s_{ij}}\left[\sum_{k=1}^C s_{ij}^{(k)}(\delta_k-1)-(s_{ij}+n_j)(\delta_{c_i}-1)+n_j(\delta_{c_j}-1)\right].
	\end{split}
\end{equation}

From the sum of \eqref{eq:first_term} and \eqref{eq:second_term}, we retrieve from simple algebra
\begin{equation}
	p_{ij}^{(c_i\rightarrow c_j)} = p_{ij}^{\mathrm{R}} \left[1+\sum_{k=1}^C s_{ij}^{(k)} (\delta_k-1)+m_i(\delta_{c_i}-1)-(m_i+s_{ij})(\delta_{c_j}-1)\right],
\end{equation}
which is equivalent to 
\begin{equation}
	\label{eq:linearized_pij}
	p_{ij}^{(c_i\rightarrow c_j)} = p_{ij}^{\mathrm{R}}\left[1+m_i(\delta_{c_i}-\delta_{c_j})+\sum_{k=1}^C s_{ij}^{(k)} (\delta_k-\delta_{c_j})\right].
\end{equation}
The linearization does not ensure that $p_{ij}$ remains between $0$ and $1$; thus, we saturate values of $p_{ij}<0$ and $p_{ij}>1$ to $0$ and $1$, respectively.

We provide a simple interpretation for the correction term in \eqref{eq:linearized_pij}. The second term in parentheses corrects for the difference in $\delta$ values between the origin and destination locations. Should the origin offer worse opportunities than the destination ($\delta_{c_i}>\delta_{c_j}$), this term increases the probability of migration or commuting from $i$ to $j$ (vice versa for $\delta_{c_i}<\delta_{c_j}$). This factor is scaled by the population at $i$ only. Thus, the term modifies the emission from $i$, such that it is independent of the population at the destination ($n_j$). 
The last term in parentheses corrects for the presence of locations of different classes between $i$ and $j$. If there are locations with better opportunities than the destination ($\delta_k<\delta_{c_j}$), it is more likely for people to stop before reaching $j$, thus reducing the probability $p_{ij}$. On the contrary, classes that are worse off than the destination ($\delta_k>\delta_{c_j}$) increase the flux from $i$ to $j$. This change is modulated by the entire population belonging to each class in a circle of radius $r_{ij}$ around $i$, excluding populations at $i$ and $j$.

\section{Selection of the value of $\delta$-s}

The linearized model in \eqref{eq:linearized_pij} offers an indication for the selection of reasonable values of $\delta$-s. We can rewrite the equation as
\begin{equation}
	p_{ij}^{(c_i\rightarrow c_j)} = p_{ij}^{\mathrm{R}} \left[1+\sum_{k=1}^C s_{ij}^{(k)} (\delta_k-1)+m_i(\delta_{c_i}-1)-(m_i+s_{ij})(\delta_{c_j}-1)\right].
\end{equation}
We find that the term $\delta_{c_i}-1$ in each equation is multiplied by the population of some location, such that in real-world datasets even a very small $\delta_{c_i}-1$ causes sizeable changes in the probabilities. Thus, these expressions offer an indication on how to choose the order of magnitude of $\delta_{c_i}-1$ and of $\delta-1$ for the case with two classes, which one could pick to be
\begin{equation}
	\delta-1\sim (0.1\div 1) \frac{1}{\max_i m_i}.
\end{equation}

\section{Results for South Sudan}

The $p$-values for the MW tests comparing the standard and modified radiation models are shown in Table \ref{tab:S1}. We find that MW tests are significant for all combinations of variables, provided $\delta$ is large enough ($\delta = 1+10^{-5}$ or $\delta = 1+10^{-4}$). For small values of $\delta$ ($1+10^{-6}$), only floods and the combination of conflicts and floods leads to significant results. In general, $p$-values for the modified radiation model based on conflicts only are larger than that of the other two variables, indicating that conflicts improve the prediction of fluxes to a lesser extent than the other two variables. This outcome may be due to the use of casualties over the 2020-2021 period only, which does not account for historical trends of violence over the previous years.
\begin{table}[htb]
	\caption{$p$-value of the MW statistical tests for the comparison between the standard and nonlinear modified radiation model with two classes in South Sudan, for each combination of variable of interest and value of $n$ in $\delta = 1+10^{n}$. Results in bold indicate significance ($\alpha = 0.05$).}
	\label{tab:S1}
	\centering
	\vspace*{6pt}
	\begin{tabular}{wr{4cm} | wc{2cm} | wc{2cm} | wc{2cm}}
		\multicolumn{1}{l|}{}   & \multicolumn{1}{c|}{$-6$} & \multicolumn{1}{c|}{$-5$} & \multicolumn{1}{c}{$-4$}  \\ \hline
		Conflicts               & $0.0505$                  & $\mathbf{<0.0001}$                  & $\mathbf{<0.0001}$         \\
		Floods                  & $\mathbf{0.0030}$                  & $\mathbf{<0.0001}$         & $\mathbf{<0.0001}$        \\
		Conflicts \& Floods & $\mathbf{0.0003}$                  & $\mathbf{<0.0001}$         & $\mathbf{<0.0001}$
	\end{tabular}
\end{table}

We show in Table \ref{tab:S2} the $z$-statistics generated from the MW test \cite{ranksum_MATLAB}. We find that the $z$-statistics is lower than the 5\% percentile of the $z$-statistics from tests with the random assignment of the class only when considering floods or both conflicts and floods, for sufficiently large values of $\delta$ ($1+10^{-5}$ and $1+10^{-4}$). At small values of $\delta$ ($1+10^{-6}$), the $z$-statistics passes the non-parametric test only for the model considering both conflicts and floods. Thus, the improvement in the prediction of fluxes when considering only conflicts (or floods at small $\delta$-s), despite providing $p$-values that are significant, may be attributed to chance (that is, a better fit of the fluxes with more parameters). On the other hand, fluxes computed accounting for floods at larger $\delta$-s and both conflicts and floods correspond to a genuine improvement of the migration model, better than what could be obtained just by adding more free parameters. 
\begin{table}[htb]
	\caption{$z$-statistics of the MW statistical tests for the comparison between the standard and nonlinear modified radiation model with two classes in South Sudan, for each combination of variable of interest and value of $n$ in $\delta = 1+10^{n}$. The table also shows the $5\%$ percentile of the empirical distribution of the $z$-statistics. Results in bold indicate that the $z$-statistics of the test with a specific variable is lower than the $5\%$ percentile of the empirical distribution of the $z$-statistics.}
	\label{tab:S2}
	\centering
	\vspace*{6pt}
	\begin{tabular}{wr{4cm} | wc{2cm} | wc{2cm} | wc{2cm}}
		\multicolumn{1}{l|}{}   & $-6$      & $-5$               & $-4$               \\ \hline
		Conflicts               & $-1.6399$ & $-4.2933$          & $-5.3214$          \\
		Floods                  & $-2.7442$ & $\mathbf{-7.0679}$          & $\mathbf{-8.6640}$          \\
		Conflicts \& Floods & $\mathbf{-3.4709}$ & $\mathbf{-9.6432}$ & $\mathbf{-11.950}$ \\ \hline
		5\% Percentile, Random  & $-2.7634$ & $-5.8068$          & $-6.9219$        
	\end{tabular}
\end{table}

Results for the statistical tests on Hamming distances are presented in Table \ref{tab:S3}. We record statistical significance for any variable of interest for any value of $\delta$ considered herein.
\begin{table}[htb]
	\caption{$p$-value of the statistical tests on Hamming distances for the comparison between the standard and nonlinear modified radiation model with two classes in South Sudan, for each combination of variable of interest and value of $n$ in $\delta = 1+10^{n}$. Results in bold indicate significance ($\alpha = 0.05$).}
	\label{tab:S3}
	\centering
	\vspace*{6pt}
	\begin{tabular}{wr{4cm} | wc{2cm} | wc{2cm} | wc{2cm}}
		& $-6$              & $-5$               & $-4$                      \\ \hline
		Conflicts               & $\mathbf{<0.0001}$          & $\mathbf{<0.0001}$  & $\mathbf{<0.0001}$         \\
		Floods                  & $\mathbf{<0.0001}$          & $\mathbf{<0.0001}$  & $\mathbf{<0.0001}$ \\
		Conflicts \& Floods & $\mathbf{<0.0001}$ & $\mathbf{<0.0001}$ & $\mathbf{<0.0001}$    
	\end{tabular}
\end{table}

Table \ref{tab:S4} shows the results of the non-parametric test on the empirical distribution of the Hamming distance. For any value of $\delta$ and variable of interest, the modified radiation model performs better than the standard one (that is, the Hamming distance is smaller). However, only when considering large enough values of $\delta$ ($1+10^{-5}$ and $1+10^{-4}$) and the effects of floods or both conflicts and floods we obtain results that are below the $5\%$ percentile of the empirical distribution of Hamming distances. At small values of $\delta$ ($1+10^{-6})$, similar to the $z$-statistics, only considering both conflicts and floods provides significant results in the non-parametric test. That is, the improvement of the model in the other conditions may only be related to the additional model parameters. Consistently with the MW tests, the modified radiation model that accounts for floods or both conflicts and floods at large enough $\delta$-s represents a genuine improvement in the description of mobility. 
\begin{table}[htb]
	\caption{Hamming distance for the standard and nonlinear modified radiation model with two classes in South Sudan, for each combination of variable of interest and value of $n$ in $\delta = 1+10^{n}$. The table also shows the $5\%$ percentile of the empirical distribution of the Hamming distances. Results in bold indicate that the Hamming distance of the model with a specific variable is lower than the $5\%$ percentile of the empirical distribution of the Hamming distances.}
	\label{tab:S4}
	\centering
	\vspace*{6pt}
	\begin{tabular}{wr{4cm} | wc{2cm} | wc{2cm} | wc{2cm}}
		Standard model: $2,718$           & $-6$    & $-5$           & $-4$           \\ \hline
		Conflicts               & $2,542$   & $2,404$          & $2,364$          \\
		Floods                  & $2,562$   & $\mathbf{2,330}$          & $\mathbf{2,266}$          \\
		Conflicts \& Floods & $\mathbf{2,466}$   & $\mathbf{2,132}$ & $\mathbf{2,034}$ \\ \hline
		5\% Percentile, Random  & $2,525$ & $2,366.5$          & $2,327.5$         
	\end{tabular}
\end{table}

\subsection{Numerical comparison of fluxes}

Figure \ref{fig:rank_South_Sudan} shows the comparison of fluxes from standard radiation model and the best performing modified radiation model (floods and conflicts, $\delta = 1+10^{-4}$) against real fluxes, in terms of their rank after sorting. While for large fluxes the rank of both models is not well-reconstructed, we find that the modified model remarkably outperforms the standard one, being much closer to the correct rank (the diagonal in Fig. \ref{fig:rank_South_Sudan}) for the majority of the domain. We interpret this result based on the hypotheses underlying our model. In fact, our model penalizes some of the locations, reducing the fluxes that reach them. We do not implement any direct effect on advantageous destination locations, such that larger fluxes may not be accurately reconstructed. On the other hand, we better reconstruct smaller fluxes that have been penalized.

\begin{figure}[htb]
	\centering
	\includegraphics[width=.5\textwidth]{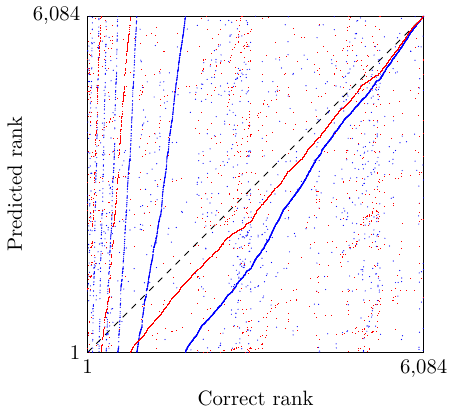} 
	\caption{Comparison of the rank of fluxes of the standard radiation model (blue, circles) and modified radiation model (red, squares) accounting for both floods and conflicts with $\delta = 1+10^{-4}$ against the rank of real fluxes. The dashed black line represents a model with zero errors.}
	\label{fig:rank_South_Sudan}
\end{figure}

\subsection{Sensitivity analysis on the penalization threshold}

We perform a sensitivity analysis by comparing the results where the $15\%$ of the counties with worst conditions based on conflicts and floods belong to the penalized class against those in which this penalization threshold is $10\%$ and $20\%$. We repeat the same tests as for the original threshold. Generation of random distributions is repeated due to the different values of the threshold, which affects the number of penalized locations. Comparisons are shown in Tables \ref{tab:sensitivity_MW_p_value}, \ref{tab:sensitivity_MW_z_value}, \ref{tab:sensitivity_Hamm_p_value}, and \ref{tab:sensitivity_Hamm_distance_values}.

We observe that the results are robust to changes in the penalization threshold, especially for larger values of $\delta$. Apart from obvious differences in numerical results, we record only a few discrepancies in the statistical significance. In the MW statistical test and corresponding non-parametric test, at the lowest value of $\delta$, we find that the modified radiation model with \textquotedblleft{}Floods\textquotedblright{} as the variable of interest outperforms the standard one when the penalization threshold is $10\%$, while none of the modified radiation models has a better performance than the standard one for a penalization threshold of $20\%$. In the Hamming distance test and corresponding non-parametric test, for a threshold of $20\%$, we barely lose significance for the modified radiation model with \textquotedblleft{}Floods\textquotedblright{} as the variable of interest.

\begin{table}[htb]
	\caption{$p$-value of the MW statistical tests for the comparison between the standard and nonlinear modified radiation model with two classes in South Sudan, for each combination of variable of interest and value of $n$ in $\delta = 1+10^{n}$. Results in bold indicate significance ($\alpha = 0.05$). Different values for the penalization threshold are used.}
	\label{tab:sensitivity_MW_p_value}
	\centering
	\vspace*{6pt}
	\begin{tabular}{wr{4cm} | wc{2cm} | wc{2cm} | wc{2cm}}
		\multicolumn{1}{r|}{\textbf{Worst $10\%$}}   & \multicolumn{1}{c|}{$-6$} & \multicolumn{1}{c|}{$-5$} & \multicolumn{1}{c}{$-4$}  \\ \hline
		Conflicts               & $0.0660$                  & $\mathbf{0.0006}$                  & $\mathbf{<0.0001}$         \\
		Floods                  & $\mathbf{0.0021}$                  & $\mathbf{<0.0001}$         & $\mathbf{<0.0001}$        \\
		Conflicts \& Floods & $\mathbf{0.0002}$                  & $\mathbf{<0.0001}$         & $\mathbf{<0.0001}$
	\end{tabular}
	
	\begin{tabular}{wr{4cm} | wc{2cm} | wc{2cm} | wc{2cm}}
		\multicolumn{1}{r|}{\textbf{Worst $15\%$}}   & \multicolumn{1}{c|}{$-6$} & \multicolumn{1}{c|}{$-5$} & \multicolumn{1}{c}{$-4$}  \\ \hline
		Conflicts               & $0.0505$                  & $\mathbf{<0.0001}$                  & $\mathbf{<0.0001}$         \\
		Floods                  & $\mathbf{0.0030}$                  & $\mathbf{<0.0001}$         & $\mathbf{<0.0001}$        \\
		Conflicts \& Floods & $\mathbf{0.0003}$                  & $\mathbf{<0.0001}$         & $\mathbf{<0.0001}$
	\end{tabular}
	
	\begin{tabular}{wr{4cm} | wc{2cm} | wc{2cm} | wc{2cm}}
		\multicolumn{1}{r|}{\textbf{Worst $20\%$}}  & \multicolumn{1}{c|}{$-6$} & \multicolumn{1}{c|}{$-5$} & \multicolumn{1}{c}{$-4$}  \\ \hline
		Conflicts               & $\mathbf{0.0046}$                  & $\mathbf{<0.0001}$                  & $\mathbf{<0.0001}$         \\
		Floods                  & $\mathbf{0.0239}$                  & $\mathbf{<0.0001}$         & $\mathbf{<0.0001}$        \\
		Conflicts \& Floods & $\mathbf{0.0010}$                  & $\mathbf{<0.0001}$         & $\mathbf{<0.0001}$
	\end{tabular}
\end{table}

\begin{table}[htb]
	\caption{$z$-statistics of the MW statistical tests for the comparison between the standard and nonlinear modified radiation model with two classes in South Sudan, for each combination of variable of interest and value of $n$ in $\delta = 1+10^{n}$. The table also shows the $5\%$ percentile of the empirical distribution of the $z$-statistics. Results in bold indicate that the $z$-statistics of the test with a specific variable is lower than the $5\%$ percentile of the empirical distribution of the $z$-statistics. Different values for the penalization threshold are used.}
	\label{tab:sensitivity_MW_z_value}
	\centering
	\vspace*{6pt}
	\begin{tabular}{wr{4cm} | wc{2cm} | wc{2cm} | wc{2cm}}
		\multicolumn{1}{r|}{\textbf{Worst $10\%$}}   & $-6$      & $-5$               & $-4$               \\ \hline
		Conflicts               & $-1.5061$ & $-3.2447$          & $-3.7636$          \\
		Floods                  & $\mathbf{-2.8558}$ & $\mathbf{-5.9966}$          & $\mathbf{-7.1053}$          \\
		Conflicts \& Floods & $\mathbf{-3.6089}$ & $\mathbf{-8.0063}$ & $\mathbf{-9.4367}$ \\ \hline
		5\% Percentile, Random  & $-2.1283$ & $-4.3821$          & $-5.2379$        
	\end{tabular}
	\begin{tabular}{wr{4cm} | wc{2cm} | wc{2cm} | wc{2cm}}
		\multicolumn{1}{r|}{\textbf{Worst $15\%$}}   & $-6$      & $-5$               & $-4$               \\ \hline
		Conflicts               & $-1.6399$ & $-4.2933$          & $-5.3214$          \\
		Floods                  & $-2.7442$ & $\mathbf{-7.0679}$          & $\mathbf{-8.6640}$          \\
		Conflicts \& Floods & $\mathbf{-3.4709}$ & $\mathbf{-9.6432}$ & $\mathbf{-11.950}$ \\ \hline
		5\% Percentile, Random  & $-2.7634$ & $-5.8068$          & $-6.9219$        
	\end{tabular}
	\begin{tabular}{wr{4cm} | wc{2cm} | wc{2cm} | wc{2cm}}
		\multicolumn{1}{r|}{\textbf{Worst $20\%$}}   & $-6$      & $-5$               & $-4$               \\ \hline
		Conflicts               & $-2.6066$ & $-6.6659$          & $-8.0270$          \\
		Floods                  & $-1.9800$ & $\mathbf{-7.3871}$          & $\mathbf{-9.1374}$          \\
		Conflicts \& Floods & $-3.0818$ & $\mathbf{-10.2565}$ & $\mathbf{-12.7559}$ \\ \hline
		5\% Percentile, Random  & $-3.4225$ & $-7.1081$          & $-8.6633$        
	\end{tabular}
\end{table}

\begin{table}[htb]
	\caption{$p$-value of the statistical tests on Hamming distances for the comparison between the standard and nonlinear modified radiation model with two classes in South Sudan, for each combination of variable of interest and value of $n$ in $\delta = 1+10^{n}$. Results in bold indicate significance ($\alpha = 0.05$). Different values for the penalization threshold are used.}
	\label{tab:sensitivity_Hamm_p_value}
	\vspace*{6pt}
	\centering
	\begin{tabular}{wr{4cm} | wc{2cm} | wc{2cm} | wc{2cm}}
		\textbf{Worst $10\%$}             & $-6$              & $-5$               & $-4$                      \\ \hline
		Conflicts               & $\mathbf{<0.0001}$          & $\mathbf{<0.0001}$  & $\mathbf{<0.0001}$         \\
		Floods                  & $\mathbf{<0.0001}$          & $\mathbf{<0.0001}$  & $\mathbf{<0.0001}$ \\
		Conflicts \& Floods & $\mathbf{<0.0001}$ & $\mathbf{<0.0001}$ & $\mathbf{<0.0001}$    
	\end{tabular}
	
	\begin{tabular}{wr{4cm} | wc{2cm} | wc{2cm} | wc{2cm}}
		\textbf{Worst $15\%$}             & $-6$              & $-5$               & $-4$                      \\ \hline
		Conflicts               & $\mathbf{<0.0001}$          & $\mathbf{<0.0001}$  & $\mathbf{<0.0001}$         \\
		Floods                  & $\mathbf{<0.0001}$          & $\mathbf{<0.0001}$  & $\mathbf{<0.0001}$ \\
		Conflicts \& Floods & $\mathbf{<0.0001}$ & $\mathbf{<0.0001}$ & $\mathbf{<0.0001}$    
	\end{tabular}
	
	\begin{tabular}{wr{4cm} | wc{2cm} | wc{2cm} | wc{2cm}}
		\textbf{Worst $20\%$}             & $-6$              & $-5$               & $-4$                      \\ \hline
		Conflicts               & $\mathbf{<0.0001}$          & $\mathbf{<0.0001}$  & $\mathbf{<0.0001}$         \\
		Floods                  & $\mathbf{<0.0001}$          & $\mathbf{<0.0001}$  & $\mathbf{<0.0001}$ \\
		Conflicts \& Floods & $\mathbf{<0.0001}$ & $\mathbf{<0.0001}$ & $\mathbf{<0.0001}$    
	\end{tabular}
	
\end{table}

\begin{table}[htb]
	\caption{Hamming distance for the standard and nonlinear modified radiation model with two classes in South Sudan, for each combination of variable of interest and value of $n$ in $\delta = 1+10^{n}$. The table also shows the $5\%$ percentile of the empirical distribution of the Hamming distances. Results in bold indicate that the Hamming distance of the model with a specific variable is lower than the $5\%$ percentile of the empirical distribution of the Hamming distances. Different values for the penalization threshold are used. The Hamming distance for the standard model is $2,718$.}
	\label{tab:sensitivity_Hamm_distance_values}
	\centering
	\vspace*{6pt}
	\begin{tabular}{wr{4cm} | wc{2cm} | wc{2cm} | wc{2cm}}
		\textbf{Worst $10\%$}      & $-6$    & $-5$           & $-4$           \\ \hline
		Conflicts               & $2,573$   & $2,486$          & $2,463$          \\
		Floods                  & $2,569$   & $\mathbf{2,401}$          & $\mathbf{2,358}$          \\
		Conflicts \& Floods & $\mathbf{2,498}$   & $\mathbf{2,269}$ & $\mathbf{2,210}$ \\ \hline
		5\% Percentile, Random  & $2,569$ & $2,467$          & $2,429.5$         
	\end{tabular}
	\begin{tabular}{wr{4cm} | wc{2cm} | wc{2cm} | wc{2cm}}
		\textbf{Worst $15\%$}      & $-6$    & $-5$           & $-4$           \\ \hline
		Conflicts               & $2,542$   & $2,404$          & $2,364$          \\
		Floods                  & $2,562$   & $\mathbf{2,330}$          & $\mathbf{2,266}$          \\
		Conflicts \& Floods & $\mathbf{2,466}$   & $\mathbf{2,132}$ & $\mathbf{2,034}$ \\ \hline
		5\% Percentile, Random  & $2,525$ & $2,366.5$          & $2,327.5$         
	\end{tabular}
	\begin{tabular}{wr{4cm} | wc{2cm} | wc{2cm} | wc{2cm}}
		\textbf{Worst $20\%$}      & $-6$    & $-5$           & $-4$           \\ \hline
		Conflicts               & $2,533$   & $2,306$          & $2,251$          \\
		Floods                  & $2,573$   & $2,280$          & $\mathbf{2,205}$          \\
		Conflicts \& Floods & $\mathbf{2,474}$   & $\mathbf{2,072}$ & $\mathbf{1,961}$ \\ \hline
		5\% Percentile, Random  & $2,481$ & $2,277$          & $2,220$         
	\end{tabular}
	
\end{table}

\subsection{Replication of statistical test with Akaike information criterion}

To further strengthen our claim that the modified radiation model is a genuine improvement of the standard radiation model, we replicate the statistical test on the Hamming distance through the Akaike information criterion, which explicitly penalizes the number of parameters used in the model \cite{konishi2008information}. To this end, we consider two likelihood functions (for each variable of interest), corresponding to the null and alternative hypotheses:
\begin{itemize}
	\item The first likelihood function assumes that the observed Hamming distances $H_\mathrm{std}$ and $H_\mathrm{mod}$ arise from a single binomial distribution with success probability $p$,
	\begin{equation}
		\mathcal{L}_1(p) = \binom{N^2}{H_\mathrm{std}}p^{H_\mathrm{std}}(1-p)^{N^2-H_\mathrm{std}}\binom{N^2}{H_\mathrm{mod}}p^{H_\mathrm{mod}}(1-p)^{N^2-H_\mathrm{mod}}.
	\end{equation}
	Note that the null hypothesis is slightly different from the one tested in the previous section, as the success probability $p$ that minimizes $\mathcal{L}_1(p)$ depends on both standard and modified radiation models.
	\item The second likelihood function assumes that the observed Hamming distances $H_\mathrm{std}$ and $H_\mathrm{mod}$ arise from two different binomial distribution with success probabilities $p_1$ and $p_2$,
	\begin{equation}
		\mathcal{L}_2(p_1,p_2) = \binom{N^2}{H_\mathrm{std}}p_1^{H_\mathrm{std}}(1-p_1)^{N^2-H_\mathrm{std}}\binom{N^2}{H_\mathrm{mod}}p_2^{H_\mathrm{mod}}(1-p_2)^{N^2-H_\mathrm{mod}}.
	\end{equation}
\end{itemize}
We then compute the Akaike information criterion (AIC) for each of these models as
\begin{equation}
	\mathrm{AIC}_i = 2k_i -2\log\left(\hat{\mathcal{L}}_i\right),
\end{equation}
where $i=1$ or $2$, $k_i$ is the number of parameters of model $i$, and $\hat{\mathcal{L}}_i$ is the maximum likelihood value for model $i$ as a function of its parameters. We assign $k_1 = 1$ ($p$ only) for the first likelihood function and $k_2 = 4$ ($p_1$, $p_2$, $\delta$, and  the percentile of the distribution of variables that separates the two classes of locations). Note that we only run the optimization for $\mathcal{L}_2$ over $p_1$ and $p_2$, such that the AIC found herein is an overestimate of the actual AIC (such that it represents a conservative estimate). Such a decision is related to the complexity of the optimization with respect to $\delta$ and the percentile of the distribution of variables that separates the two classes of locations, which are selected empirically.

Finally, we compute the relative likelihood of the two models as
\begin{equation}
	\mathrm{RL}_i = \mathrm{exp}\left(\frac{\min_j\mathrm{AIC}_j-\mathrm{AIC}_i}{2}\right),
\end{equation}
which is proportional to the probability that model $i$ minimizes information loss. The model that most probably minimizes information loss will have $\mathrm{RL}=1$, while the other will have $\mathrm{RL}<1$. A smaller value of the relative likelihood indicates a model that is more unlikely to correctly represent the information contained in the data.

Results for the relative likelihoods are shown in Tab. \ref{tab:S5}.  
For large enough values of $\delta$ ($1+10^{-5}$ and $1+10^{-4}$), all combinations of variables yield a model that is at least 10,000 times more likely to minimize information loss compared to the standard radiation model. Interestingly, the modified radiation model that only considers conflicts is over $10^7$ times more likely than the standard one to minimize information loss at large values of $\delta$ ($1+10^{-4}$), in contrast with previous results, suggesting that the non-parametric test may be conservative. These results further confirm that the improvement in the predictions in the modified radiation model is not only related to the additional number of parameters used.
\begin{table}[htb]
	\caption{Relative likelihood for the standard ($\mathrm{RL}_1$) and modified radiation model ($\mathrm{RL}_2$), for the Hamming distance test in South Sudan, with different variables used for the selection of classes and values of $n$ in $\delta = 1+10^{n}$. Results in bold indicate instances in which the standard radiation model is less than 5\% likely to minimize information loss compared to the modified radiation model.}
	\label{tab:S5}
	\centering
	\vspace*{6pt}
	\begin{tabular}{rr|c|c|c}
		\multicolumn{1}{l}{}                                          & \multicolumn{1}{l|}{} & $-6$     & $-5$              & $-4$                            \\ \hline
		\multicolumn{1}{r|}{\multirow{2}{*}{Conflicts}}               & $\mathrm{RL}_1$       & $0.1122$      & $\mathbf{1.2042\times 10^{-6}}$               & $\mathbf{1.2650\times 10^{-8}}$                             \\
		\multicolumn{1}{r|}{}                                         & $\mathrm{RL}_2$       & $1$ & $1$          & $1$                        \\ \hline
		\multicolumn{1}{r|}{\multirow{2}{*}{Floods}}                  & $\mathrm{RL}_1$       & $0.3425$      & $\mathbf{1.6866\times 10^{-10}}$          & $\mathbf{1.6213\times 10^{-14}}$               \\
		\multicolumn{1}{r|}{}                                         & $\mathrm{RL}_2$       & $1$ & $1$               & $1$                             \\ \hline
		\multicolumn{1}{r|}{\multirow{2}{*}{Conflicts \& Floods}} & $\mathrm{RL}_1$       & $\mathbf{4.6439\times 10^{-4}}$      & $\mathbf{4.8977\times 10^{-25}}$ & $\mathbf{1.3507\times 10^{-34}}$ \\
		\multicolumn{1}{r|}{}                                         & $\mathrm{RL}_2$       & $1$ & $1$               & $1$                            
	\end{tabular}
\end{table}

\section{Plots of variables of interests in the United States}

Figure \ref{fig:S3} shows the counties in the US that have a value of each variable of interest (Gini index, poverty ratio, ratio between the median rent and median household income, and unemployment rate) over the $90\%$ percentile.

\begin{figure}[htbp]
	\centering
	\begin{tabular}{c}
		(a) \\
		\includegraphics[width=0.5\textwidth]{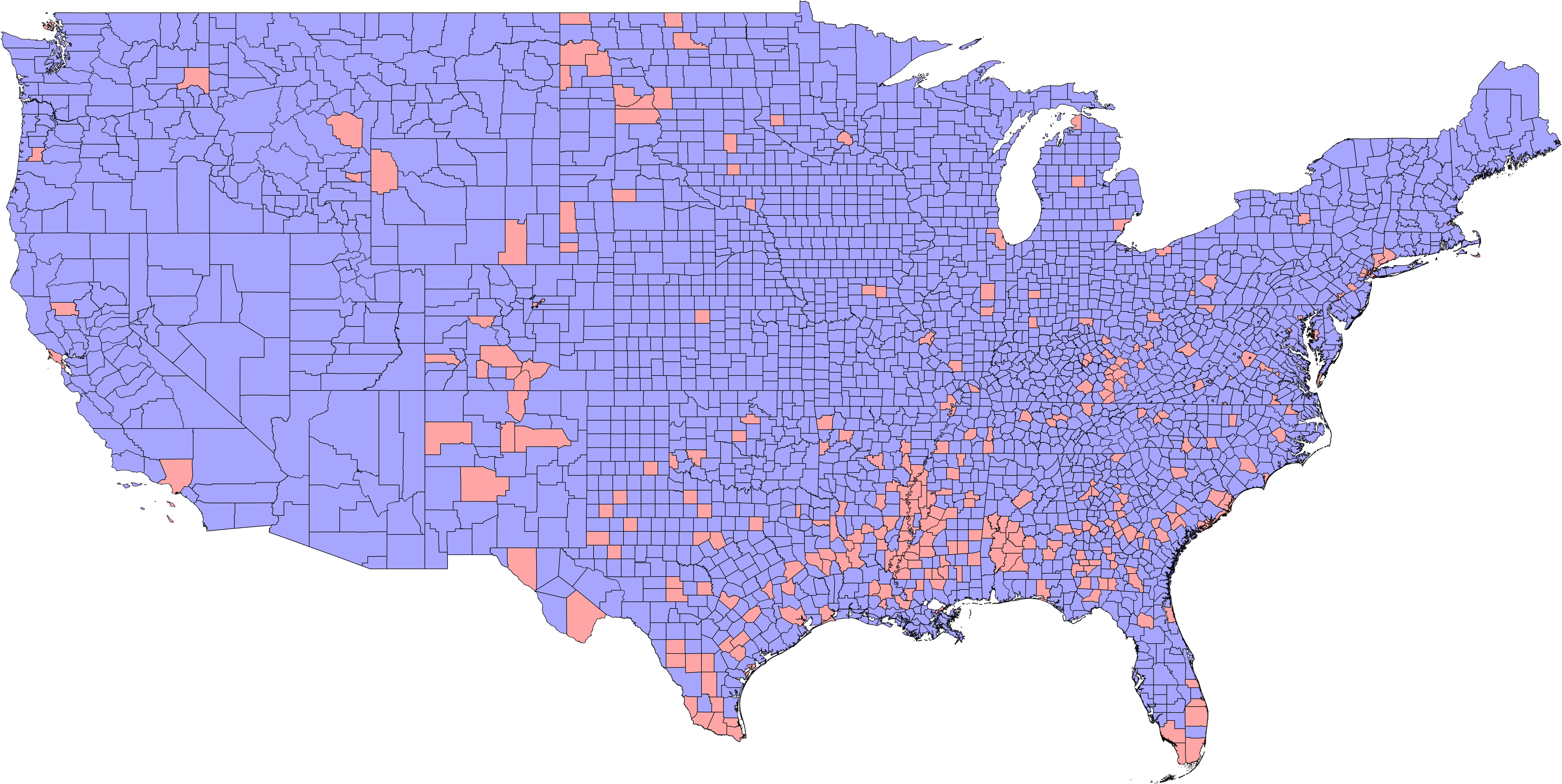} \\
		(b) \\
		\includegraphics[width=0.5\textwidth]{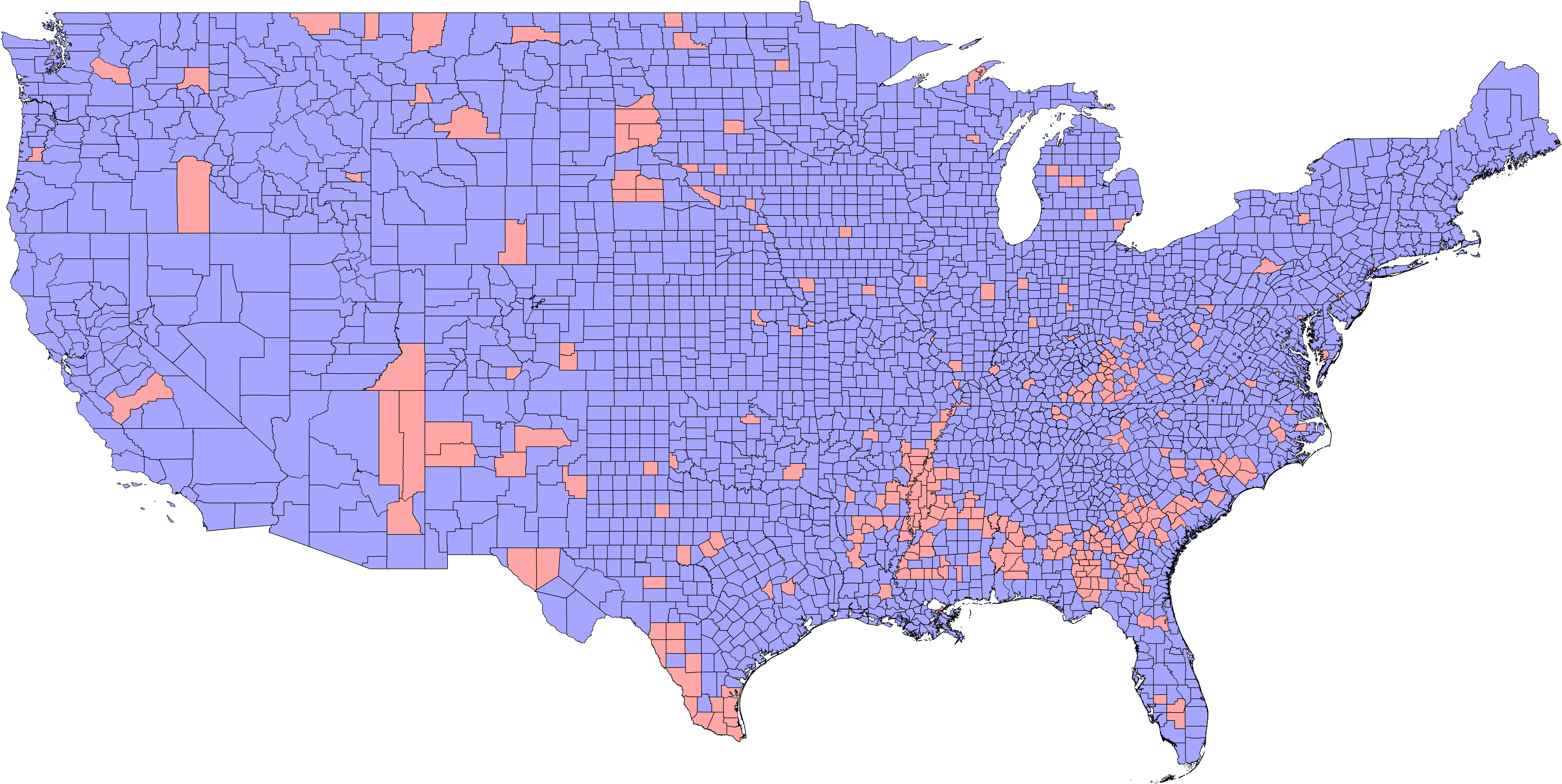} \\
		(c) \\
		\includegraphics[width=0.5\textwidth]{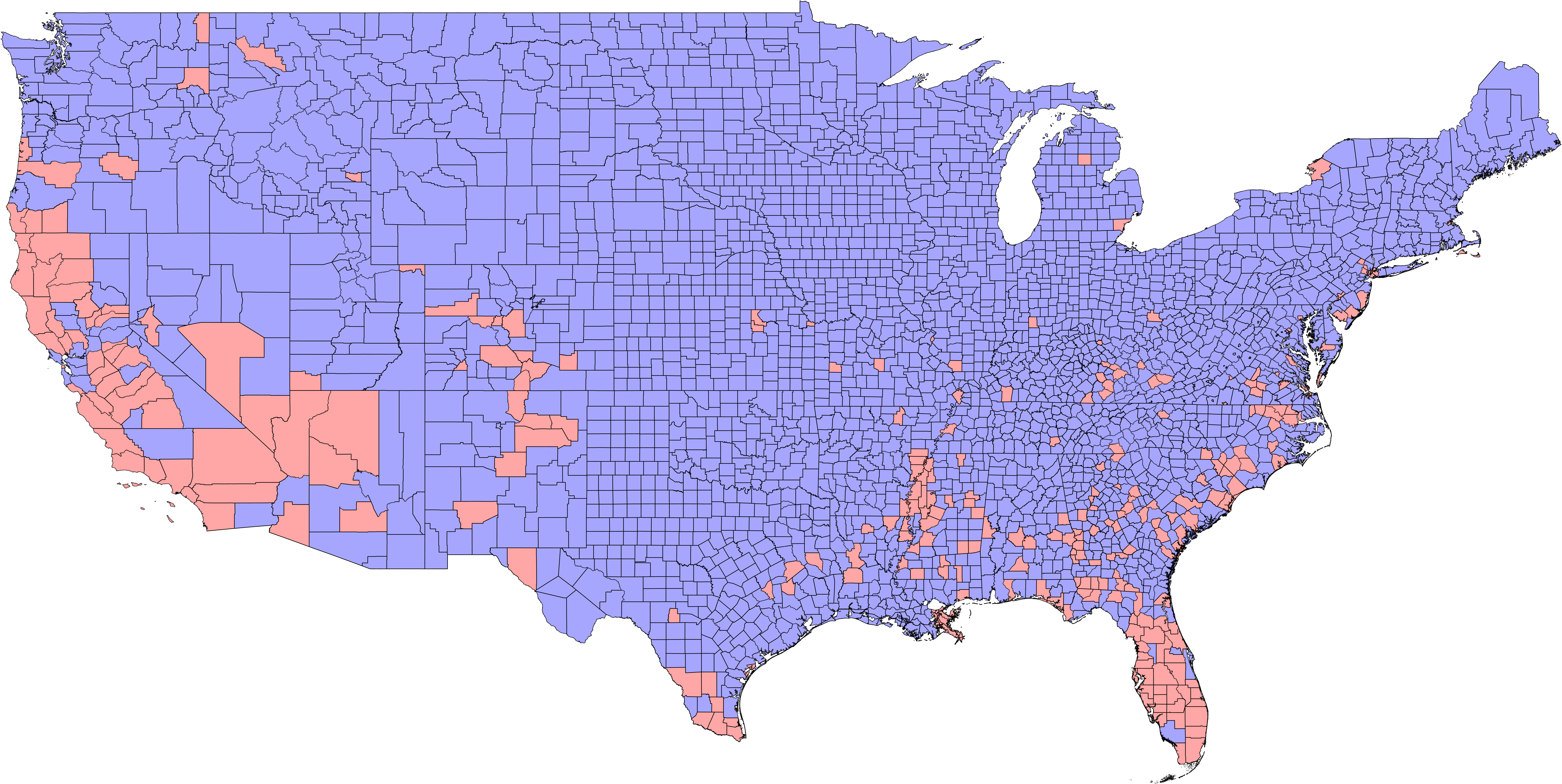} \\
		(d) \\
		\includegraphics[width=0.5\textwidth]{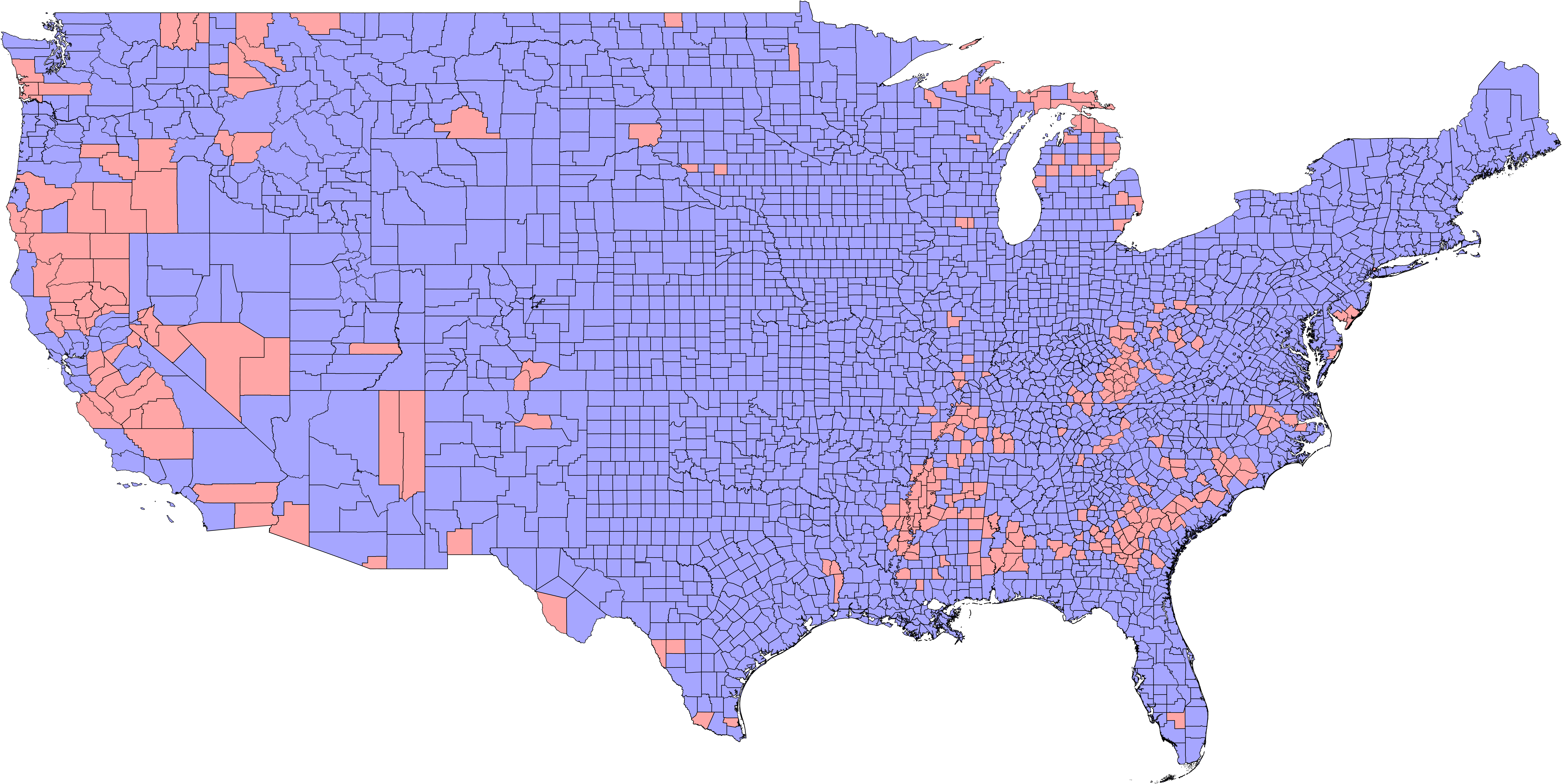}
	\end{tabular}
	\caption{US counties (red) with the largest 10$\%$ values of: (a) Gini index; (b) poverty ratio; (c) ratio between the median rent and median household income; and (d) unemployment rate.}
	\label{fig:S3}
\end{figure}

\section{Results for the United States}

We show the $p$-values for the MW tests comparing the standard and modified radiation models in Table \ref{tab:S6}. For moderate values of $\Delta$ ($10^{-7}$), all variables of interests used in the modified radiation models yield statistical significance. For smaller values of $\Delta$ ($10^{-8}$), we only register a significant result for the ratio between median rent and median income. At higher $\Delta$ values, only modified radiation models based on poverty ratio and unemployment rate outperform the standard radiation model. We interpret the lack of significance at small values of $\Delta$ as an effect of the socioeconomic variables which is too weak to generate substantial changes in the fluxes. On the other hand, for larger values of $\Delta$, the rejection of higher-order terms in the linearized model is likely to generate large errors (note that $\delta$ values are multiplied by large coefficients, so that even small variations of $\delta$ from $1$ may cause large errors compared to the nonlinear model).
\begin{table}[htb]
	\caption{$p$-value of the MW statistical tests for the comparison between the standard and linearized modified radiation model with arbitrary number of classes in the US, for each combination of variable of interest and value of $n$ in $\Delta = 10^{n}$. Results in bold indicate significance ($\alpha = 0.05$).}
	\label{tab:S6}
	\centering
	\vspace*{6pt}
	\begin{tabular}{wr{4cm} | wc{2cm} | wc{2cm} | wc{2cm}}
		\multicolumn{1}{l|}{} & $-8$              & $-7$                  & $-6$                  \\ \hline
		Gini index                    & $0.1045$          & $\mathbf{<0.0001}$    & $0.8683$              \\
		Poverty ratio                    & $0.0652$          & $\mathbf{<0.0001}$ & $\mathbf{<0.0001}$ \\
		Median rent/household income                     & $\mathbf{0.0288}$ & $\mathbf{<0.0001}$ & $1$ \\
		Unemployment rate                    & $0.2358$          & $\mathbf{<0.0001}$ & $\mathbf{<0.0001}$                  
	\end{tabular}
\end{table}

Results for the $z$-statistics of the MW tests are shown in Table \ref{tab:S7}. For small values of $\Delta$ ($10^{-8}$), the $z$-statistics for all variables is lower than the $5\%$ percentile of the empirical distribution. Similarly, at moderate values of $\Delta$ ($10^{-7}$), we find results below the $5\%$ percentile for all variables but the unemployment rate. In these conditions, the proposed model genuinely outperforms the standard radiation model. On the other hand, at larger values of $\Delta$ ($10^{-6}$), all improvements can be traced back to the larger number of parameters in the model.
\begin{table}[htb]
	\caption{$z$-statistics of the MW statistical tests for the comparison between the standard and linearized modified radiation model with arbitrary number of classes in the US, for each combination of variable of interest and value of $n$ in $\Delta = 10^{n}$. The table also shows the $5\%$ percentile of the empirical distribution of the $z$-statistics. Results in bold indicate that the $z$-statistics of the test with a specific variable is lower than the $5\%$ percentile of the empirical distribution of the $z$-statistics.}
	\label{tab:S7}
	\centering
	\vspace*{6pt}
	\begin{tabular}{wr{4cm} | wc{2cm} | wc{2cm} | wc{2cm}}
		\multicolumn{1}{l|}{} & $-8$              & $-7$                  & $-6$                  \\ \hline
		Gini index                    & $\mathbf{-1.2562}$          & $\mathbf{-6.4471}$    & $1.1187$              \\
		Poverty ratio                    & $\mathbf{-1.5127}$          & $\mathbf{-7.8541}$ & $-15.555$ \\
		Median rent/household income                     & $\mathbf{-1.8990}$ & $\mathbf{-6.7372}$ & $5.6317$ \\
		Unemployment rate                    & $\mathbf{-0.7200}$          & $-4.2564$ & $-11.871$             \\ \hline
		5\% Percentile, Random  & $-0.6783$ & $-5.9074$          & $-16.042$     
	\end{tabular}
\end{table}

Table \ref{tab:S8} shows the $p$-values of the statistical tests on Hamming distances for US commuting patterns, comparing standard and linearized modified radiation models.  Regardless of the value of $\Delta$, we register statistically significant differences between the models, apart for a single condition (unemployment ratio with $\Delta = 10^{-8}$). \textit{}Results on the non-parametric test on Hamming distances are displayed in Table \ref{tab:S9}. The Hamming distance for the standard radiation model is higher than that for any modified radiation model. At low and moderate values of $\Delta$ ($10^{-8}$ and $10^{-7}$), we find that the Hamming distance is below the $5\%$ percentile of the empirical distribution for any variable (apart for unemployment ratio with $\Delta = 10^{-7}$). Therefore, in all of these conditions, the modified radiation model performs better than just fitting with additional model parameters. For larger values of $\Delta$ ($10^{-6}$), only the modified radiation model with poverty ratio shows a Hamming distance below the $5\%$ percentile of the empirical distribution. In the cases in which the non-parametric test fails, we attribute the improvements in the predictions of fluxes to the additional model parameters.
\begin{table}[htb]
	\caption{$p$-value of the statistical tests on Hamming distances for the comparison between the standard and linearized modified radiation model with arbitrary number of classes in the US, for each combination of variable of interest and value of $n$ in $\Delta = 10^{n}$. Results in bold indicate significance ($\alpha = 0.05$).}
	\label{tab:S8}
	\centering
	\vspace*{6pt}
	\begin{tabular}{wr{4cm} | wc{2cm} | wc{2cm} | wc{2cm}}
		\multicolumn{1}{l|}{} & $-8$              & $-7$                  & $-6$                  \\ \hline
		Gini index                    & $\mathbf{0.0278}$          & $\mathbf{<0.0001}$    & $\mathbf{<0.0001}$              \\
		Poverty ratio                    & $\mathbf{0.0247}$          & $\mathbf{<0.0001}$ & $\mathbf{<0.0001}$ \\
		Median rent/household income                     & $\mathbf{0.0035}$ & $\mathbf{<0.0001}$ & $\mathbf{<0.0001}$ \\
		Unemployment rate                    & $0.1641$          & $\mathbf{<0.0001}$ & $\mathbf{<0.0001}$                  
	\end{tabular}
\end{table}
\begin{table}[htb]
	\caption{Hamming distance for the standard and linearized modified radiation model with arbitrary number of classes in the US, for each combination of variable of interest and value of $n$ in $\Delta = 10^{n}$. The table also shows the $5\%$ percentile of the empirical distribution of Hamming distances. Results in bold indicate that the Hamming distance of the model with a specific variable is lower than the $5\%$ percentile of the empirical distribution of the Hamming distances.}
	\label{tab:S9}
	\centering
	\vspace*{6pt}
	\begin{tabular}{wr{4cm} | wc{2cm} | wc{2cm} | wc{2cm}}
		Standard model: $341,853$ & $-8$              & $-7$                  & $-6$                  \\ \hline
		Gini index                    & $\mathbf{340,754}$          & $\mathbf{336,052}$    & $337,443$              \\
		Poverty ratio                    & $\mathbf{340,724}$          & $\mathbf{336,051}$ & $\mathbf{330,436}$ \\
		Median rent/household income                     & $\mathbf{340,306}$ & $\mathbf{335,676}$ & $338,684$ \\
		Unemployment rate                    & $\mathbf{341,291}$          & $338,746$ & $333,428$             \\ \hline
		5\% Percentile, Random  & $341,395$ & $338,365$          & $330,866$     
	\end{tabular}
\end{table}
\end{widetext}

\end{document}